# Coupling $q$-deformed dark energy to dark matter

**Emre DİL**

Department of Physics, Sinop University, 57000, Korucuk, Sinop-TURKEY
e-mail: emredil@sakarya.edu.tr

**Abstract:** We propose a novel coupled dark energy model which is assumed to occur as a q-deformed scalar field and investigate whether it will provide an expanding universe phase. We consider the q-deformed dark energy as coupled to dark matter inhomogeneities. We perform the phase-space analysis of the model by numerical methods and find the late-time accelerated attractor solutions. The attractor solutions imply that the coupled $q$-deformed dark energy model is consistent with the conventional dark energy models satisfying an acceleration phase of universe. At the end, we compare the cosmological parameters of deformed and standard dark energy models and interpret the implications.

## 1. Introduction

The standard model of cosmology states that approximately 5% of the energy content of the universe belongs to ordinary baryonic matter of standard model of particle physics. The other 95% of the energy content of the universe is made of the dark sector. Particularly, 25% of the content is an unknown form of matter having a mass but in non-baryonic form that is called as dark matter. Remaining 70% of the content consists of an unknown form of energy named as dark energy. On the other hand, it is known that the universe is experiencing an accelerating expansion by astrophysical observations, Supernova Ia, large scale structure, the baryon acoustic oscillations and cosmic microwave background radiation. The dark energy is assumed to responsible from the late-time accelerated expansion of the universe. The dark energy component of the universe is not clustered but spread all over the universe and it generates a gravitational repulsion due to its negative pressure for driving the acceleration of the expansion of the universe [1-10].

In order to explain the viable mechanism for the accelerated expansion of universe, cosmologists have proposed various dynamical models for dark energy and dark matter, which include possible interactions between dark energy, dark matter and other fields, such as gravitation. Especially, the coupling between dark energy and dark matter is proposed since the energy densities of two dark components are of same order in magnitude today [11-18].

Since there is a great number of candidates for the constitution of the dark energy, such as cosmological constant, quintessence, phantom and tachyon fields, different interactions have been proposed between these constituents, dark matter and gravitational field in the framework of general relativity [19-29]. However, the corresponding dynamical analysis of the interactions between different dark energy models and the dark matter and the gravity have been studied in the framework of teleparallel gravity which uses the torsion tensor instead of the curvature tensor of general relativity [30-39].

The main motivation of this study comes from the recent studies in the literature [40-45] which involves the deformation of the standard scalar field equations representing the dark energy. In this study we propose a novel dark energy model as a *q*-deformed scalar field interacting with the dark matter. Since the dark energy is a negative-pressure scalar field, this scalar field can be considered as a *q*-deformed scalar field. Because the *q*-deformed scalar field is in fact a *q*-deformed boson model, and the statistical mechanical studies of *q*-deformed boson models have shown that the pressure of the deformed bosons is generally negative. Not only the different deformed boson models have a negative pressure, but also the different deformed fermion models can take negative pressure values [46-50]. Here, we consider the *q*-deformed bosons and propose that the scalar field which is produced by these deformed bosons constitutes the dark energy in the universe. We also investigate the dynamics of the coupling *q*-deformed dark energy and dark matter inhomogeneities in the Friedmann-Robertson-Walker (FRW) space-time. In order to confirm our proposal, we perform the phase space analysis of the model whether the late-time stable attractor solutions exist, since the stable attractor solutions imply the accelerating expansion of the universe. We finally compare the cosmological parameters of the *q*-deformed and standard dark energy model and interpret the implications of the comparison.

## 2. Dynamics of the model: Coupling *q*-deformed dark energy to dark matter

In our model, the dark energy consists of the scalar field whose field equations are defined by the *q*-deformed boson fields. Since the idea of *q*-deformation to the single particle quantum mechanics is a previous establishment in the literature [51-53], it has been natural to construct a *q*-deformed quantum field theory [54-56]. While the bosonic counterpart of the deformed particle fields corresponds to the deformed scalar field, the fermionic one corresponds to the deformed vector field. Here we take into account the $q$-deformed boson field as the $q$-deformed scalar field which constitutes the dark energy under consideration. The $q$-deformed dark energy couples to dark matter inhomogeneities in our model. Now we begin with the defining the $q$-deformed dark energy in the FRW geometry.

Quantum field theory in curved space-time is important on the understanding of the scenario in the Early Universe. Quantum mechanically, constructing the coherent states for any mode of the scalar field translates the behavior of the classical scalar field around the initial singularity into quantum field regime. At the present universe, the quantum mechanical state of the scalar field around the initial singularity cannot be determined by an observer. Therefore, Hawking states that this undeterministic nature can be described by taking the random superposition of all possible states in that space-time. Berger has realized this by taking the superposition of coherent states randomly. Parker has studied the particle creation in the expanding universe with a non-quantized gravitational metric. When the evolution of the scalar field is considered in an expanding universe, Goodison and Tom has stated that if the field quanta obey the Bose or Fermi statistics, then the particle creation does not occur in the vacuum state. Therefore, the scalar field dark energy must be described in terms of the deformed bosons or fermions in the coherent states or squeezed state [56-62]. Also the $q$-deformed dark energy is a generalization of the standard scalar field dark energy. The free parameter $q$ makes it possible to obtain a desired dark energy model with a preferred interaction or coupling by setting up the deformation parameter to the suitable value.

This motivates us to describe the dark energy as a $q$-deformed scalar field interacting with the dark matter. We can give the Dirac-Born-Infeld type action of the model as $S=\kappa^2\int(\ell_g+\ell_{\phi_q}+\ell_m+\ell_r)\,d^4x$, where $\kappa^2=8\pi G$, $\ell_g=-(1/2\kappa)\,g\,R$, $\ell_m$ and $\ell_r$ are the gravitational, dark matter and radiation Lagrangian densities. Then the deformed dark energy Lagrangian density is given for $(+,-,-,-)$ metric signature as [63]

$$\ell_{\phi_q}=g\frac{1}{2}\left[g^{\mu\nu}(\nabla_\mu\phi_q)(\nabla_\nu\phi_q)-m^2{\phi_q}^2\right], \tag{1}$$

where $g=\sqrt{-\det g_{\mu\nu}}$ and $\phi_q$ is the $q$-deformed scalar field operator for the dark energy, and $\nabla_\mu$ is the covariant derivative which is in fact the ordinary partial derivative $\partial_\mu$ for the scalar field. From the variation of the $q$-deformed scalar field Lagrangian density with respect to the deformed field, we obtain the deformed Klein-Gordon equation:

$$\frac{\partial \ell_{\phi_q}}{\partial \phi_q} - \partial_\mu \left( \frac{\partial \ell_{\phi_q}}{\partial (\partial_\mu \phi_q)} \right) = 0$$

$$(\partial_\mu \partial^\mu)\phi_q + m^2 \phi_q = 0 \,. \tag{2}$$

Also we get the energy-momentum tensor of the scalar field dark energy from the variation of Lagrangian $\ell_{\phi_q}$ with respect to the metric tensor, such as

$$\begin{aligned} T_{\mu\nu}^{\phi_q} &= \frac{2}{g}\left( \frac{\partial \ell_{\phi_q}}{\partial g^{\mu\nu}} - g_{\mu\nu}\ell_{\phi_q} \right) \\ &= \partial_\mu \phi_q \, \partial_\nu \phi_q - \frac{1}{2} g_{\mu\nu}(g^{\alpha\beta}\partial_\alpha \phi_q \, \partial_\beta \phi_q - m^2 {\phi_q}^2) \end{aligned} , \tag{3}$$

from which the timelike and spacelike parts of $T_{\mu\nu}^{\phi_q}$ reads as follows,

$$T_{00}^{\phi_q} = \frac{1}{2}\dot{\phi}_q{}^2 - \frac{1}{2} g^{ii}(\partial_i \phi_q)^2 + \frac{1}{2} m^2 {\phi_q}^2 \,, \tag{4}$$

$$T_{ii}^{\phi_q} = -\frac{1}{2} g_{ii}\dot{\phi}_q{}^2 + \frac{1}{2}(\partial_i \phi_q)^2 - \frac{1}{2} g_{ii} g^{jj}(\partial_j \phi_q)^2 + \frac{1}{2} g_{ii} m^2 {\phi_q}^2 \,, \tag{5}$$

where $i, j = 1, 2, 3$ represents the spacelike components.

In order to calculate the deformed energy density and the pressure functions of the $q$-deformed energy from the energy-momentum components (4) and (5), we need to consider the quantum field theoretical description of the deformed scalar field in a FRW background with the metric

$$ds^2 = dt^2 - a^2(t)[dx^2 + dy^2 + dz^2] \,. \tag{6}$$

Then the canonically quantized $q$-deformed scalar field $\phi_q$ is introduced in terms of the Fourier expansion [57]

$$\phi_q = \int d^3k\, [a_q(k) f_k + a_q^+(k) f_k^*], \tag{7}$$

where $a_q(k)$ and $a_q^+(k)$ are the $q$-deformed boson annihilation and creation operators for the quanta of the $q$-deformed scalar field dark energy in the $k$-th mode. $k$ denotes the spatial wave vector and obeys the relativistic energy conservation law $\omega^2 = -g^{ii}k^2 + m^2$. Here $f_k$ is a set of orthonormal mode solutions of the deformed Klein-Gordon equation, such as

$$f_k(x^\mu) = \frac{\exp(ik_\mu x^\mu)}{\sqrt{(2\pi)^3 2\omega}}, \tag{8}$$

where $k_\mu = (\omega, -k)$ is the momentum four vector and satisfy the relations

$$\int d^3x\, f_k f_{k'}^* = \frac{\delta^3(-k+k')}{2\omega}, \quad \int d^3x\, f_k f_{k'} = \frac{e^{2i\omega t}\delta^3(-k-k')}{2\omega},$$
$$\int d^3x\, f_k^* f_{k'} = \frac{\delta^3(k-k')}{2\omega}, \quad \int d^3x\, f_k^* f_{k'}^* = \frac{e^{-2i\omega t}\delta^3(k+k')}{2\omega}. \tag{9}$$

We can give the spatial average of the $q$-deformed scalar field energy-momentum tensor $T_{\mu\nu}^{\phi_q}$ as [59]

$$\overline{T}_{\mu\nu}^{\phi_q} = \int d^3x T_{\mu\nu}^{\phi_q}. \tag{10}$$

By using (10), the spatial average of the timelike energy-momentum tensor component is obtained as

$$\overline{T}_{00}^{\phi_q} = \int d^3x \left[\frac{1}{2}\dot{\phi}_q^{\,2} - \frac{1}{2} g^{ii}(\partial_i \phi_q)^2 + \frac{1}{2} m^2 \phi_q^{\,2}\right]. \tag{11}$$

This can be determined from the equations (7)-(9) term by term, as follows

$$\begin{aligned}&\frac{1}{2}m^2\int d^3x\,\phi_q{}^2\\&=\frac{1}{2}m^2\int d^3x\iint d^3k\,d^3k'[a_q(k)f_k+a_q^+(k)f_k^*][a_q(k')f_{k'}+a_q^+(k')f_{k'}^*]\\&=\frac{1}{2}\int d^3k\frac{m^2}{2\omega}[a_q(k)a_q(-k)e^{2i\omega t}+a_q^+(k)a_q(k)+a_q(k)a_q^+(k)+a_q^+(k)a_q^+(-k)e^{-2i\omega t}]\end{aligned}\quad, \tag{12}$$

and,

$$\begin{aligned}&\frac{1}{2}\int d^3x\,\dot{\phi}_q{}^2\\&=\frac{1}{2}\int d^3k\frac{\omega}{2}[-a_q(k)a_q(-k)e^{2i\omega t}+a_q^+(k)a_q(k)+a_q(k)a_q^+(k)-a_q^+(k)a_q^+(-k)e^{-2i\omega t}]\end{aligned}\;, \tag{13}$$

and

$$\begin{aligned}&\frac{1}{2}g^{ii}\int d^3x(\partial_i\,\phi_q)^2\\&=\frac{1}{2}\int d^3k\frac{g^{ii}k^2}{2\omega}[a_q(k)a_q(-k)e^{2i\omega t}+a_q^+(k)a_q(k)+a_q(k)a_q^+(k)+a_q^+(k)a_q^+(-k)e^{-2i\omega t}]\end{aligned}\;. \tag{14}$$

Combining (12), (13) and (14) gives

$$\overline{T}_{00}^{\phi_q}=\frac{1}{2}\int d^3k\,\omega[a_q(k)a_q^+(k)+a_q^+(k)a_q(k)], \tag{15}$$

where $\omega^2=m^2-k^2/a^2(t)$ for the FRW spacetime. Correspondingly, the average of the spacelike energy-momentum tensor component can be determined, such as

$$
\begin{aligned}
\overline{T}_{ii}^{\phi_q} &= \int d^3x \left[ -\frac{1}{2} g_{ii} \dot{\phi}_q^{\,2} + \frac{1}{2} (\partial_i \phi_q)^2 - \frac{1}{2} g_{ii} g^{jj} (\partial_j \phi_q)^2 + \frac{1}{2} g_{ii} m^2 \phi_q^{\,2} \right] \\
&= -\frac{1}{2} \int d^3k \frac{g_{ii}\,\omega}{2} [-a_q(k) a_q(-k) e^{2i\omega t} + a_q^+(k) a_q(k) + a_q(k) a_q^+(k) - a_q^+(k) a_q^+(-k) e^{-2i\omega t}] \\
&+ \int d^3k \frac{k_i^2}{2\omega} [a_q(k) a_q(-k) e^{2i\omega t} + a_q^+(k) a_q(k) + a_q(k) a_q^+(k) + a_q^+(k) a_q^+(-k) e^{-2i\omega t}] \\
&+ \frac{1}{2} \int d^3k \frac{g_{ii} g^{jj} k^2}{2\omega} [a_q(k) a_q(-k) e^{2i\omega t} + a_q^+(k) a_q(k) + a_q(k) a_q^+(k) + a_q^+(k) a_q^+(-k) e^{-2i\omega t}] \\
&+ \frac{1}{2} \int d^3k \frac{g_{ii} m^2}{2\omega} [a_q(k) a_q(-k) e^{2i\omega t} + a_q^+(k) a_q(k) + a_q(k) a_q^+(k) + a_q^+(k) a_q^+(-k) e^{-2i\omega t}]
\end{aligned}
\quad . \tag{16}
$$

From the identities $a_q(-k) e^{2i\omega t} = a_q^+(k)$ and $a_q^+(-k) e^{-2i\omega t} = a_q(k)$ , (16) turns out to be

$$
\overline{T}_{ii}^{\phi_q} = \frac{1}{2} \int d^3k \frac{1}{\omega} [2k_i^2 - a^2(t)\omega^2][a_q(k) a_q^+(k) + a_q^+(k) a_q(k)], \tag{17}
$$

for the FRW geometry, and $k_i$ is the $i$-th spatial component of the wave vector.

The $q$-deformation of a quantum field theory is constructed from the standard algebra satisfied by the annihilation and creation operators introduced in the canonical quantization of the field. The deformation of a standard boson algebra satisfied by the annihilation and creation operators of a bosonic quantum field theory, is firstly realized by Arik-Coon [51], and then Macfarlane and Biedenharn [52,53] independently realized the deformation of boson algebra different from Arık-Coon. Hence, the $q$-deformed bosonic quantum field theory of the scalar field dark energy is constructed by the $q$-deformed algebra of the operators $a_q(k)$ and $a_q^+(k)$ in (7), such that

$$
a_q(k) a_q^+(k') - q^2 a_q^+(k') a_q(k) = \delta^3(k - k'), \tag{18}
$$

$$
a_q(k) a_q(k') - q^2 a_q(k') a_q(k) = 0, \tag{19}
$$

$$
[N_k] = a_q^+(k) a_q(k). \tag{20}
$$

Here, $q$ is a real deformation parameter, and $[N_k]$ is the deformed number operator whose eigenvalue spectrum is given as [51]

$$[n_k] = \frac{1-q^{2n_k}}{1-q^2}, \tag{21}$$

where $n_k$ is the eigenvalue of the standard number operator $N_k$.

The corresponding vector space of the annihilation and creation operators for the $q$-deformed scalar field dark energy are the $q$-deformed Fock space state vectors, which gives information about the number of particles in the corresponding state. The $q$-deformed bosonic annihilation, creation operators $a_q(k)$ and $a_q^+(k)$ act on the Fock states $\left|n_k\right\rangle$ as

$$\begin{aligned} a_q(k)\left|n_k\right\rangle &= \sqrt{[n_k]}\left|n_k - 1\right\rangle, \\ a_q^+(k)\left|n_k\right\rangle &= \sqrt{[n_k + 1]}\left|n_k + 1\right\rangle, \\ [N_k]\left|n_k\right\rangle = a_q^+(k)\, a_q(k)\left|n_k\right\rangle &= [n_k]\left|n_k\right\rangle. \end{aligned} \tag{22}$$

By taking the quantum expectation values of the spatial averages of energy-momentum tensor with respect to the Fock basis $\left|n_k\right\rangle$, we obtain the energy density and the pressure of the $q$-deformed dark energy. Using $\rho_{\phi_q} = T_0^0$ and $p_{\phi_q} = -T_i^i$ for the energy density and pressure of the $q$-deformed scalar field dark energy, we obtain,

$$\begin{aligned} \int \rho_{\phi_q} d^3x &= \left\langle n_k \left| \overline{T}_0^{0\phi_q} \right| n_k \right\rangle \\ &= \int d^3k \frac{\omega}{2} \left\langle n_k \left| [a_q(k)\, a_q^+(k) + a_q^+(k)\, a_q(k)] \right| n_k \right\rangle, \\ &= \left((1+q^2)[n_k]+1\right) \int d^3k_\phi \frac{\omega_\phi}{2} \end{aligned} \tag{23}$$

$q$-deformed boson algebra in (18) is used in the second line. Because the $q$-deformed boson algebra in (18)-(20) transforms to be the standard boson algebra and $[n_k] = n_k$ in the $q \to 1$ limit, the energy density $\rho_{\phi_q}$ of the *q*-deformed dark energy transforms into the energy density $\rho_\phi$ of the standard dark energy as

$$\int \rho_\phi d^3x = (2n_k + 1)\int d^3k_\phi \, \frac{\omega_\phi}{2} \, . \tag{24}$$

Hence, the energy density $\rho_{\phi_q}$ of the *q*-deformed dark energy can be written in terms of the energy density $\rho_\phi$ of the standard dark energy by

$$\rho_{\phi_q} = \frac{(1+q^2)[n_k]+1}{2n_k+1}\rho_\phi = \Delta_q(n_k)\rho_\phi \, . \tag{25}$$

Accordingly, the pressure of the *q*-deformed scalar field dark energy can be written from (7), such as

$$\begin{aligned}\int p_{\phi_q} d^3x &= \left\langle n_k \left| -\overline{T}_i^{\,i\phi_q} \right| n_k \right\rangle \\ &= \int d^3k \frac{1}{2\omega}[\frac{2k^2}{a^2(t)} - \omega^2]\left\langle n_k \left| [a_q(k)a_q^+(k) + a_q^+(k)a_q(k)] \right| n_k \right\rangle , \\ &= \left((1+q^2)[n_k]+1\right)\int d^3k \frac{1}{2\omega}[\frac{2k^2}{a^2(t)} - \omega^2]\end{aligned} \tag{26}$$

where $g^{ii}k_i^2 = (1/a^2(t))(k_1^2 + k_2^2 + k_3^2) = k^2/a^2(t)$ is used. In the $q \to 1$ limit, the *q*-deformed pressure $p_{\phi_q}$ of dark energy transforms to the standard pressure $p_\phi$ of the dark energy, such that

$$\int p_\phi d^3x = (2n_k + 1)\int d^3k \frac{1}{2\omega}[\frac{2k^2}{a^2(t)} - \omega^2] . \tag{27}$$

Consequently, the *q*-deformed pressure $p_{\phi_q}$ of dark energy can be obtained in terms of the standard pressure $p_\phi$ of the dark energy, thus

$$p_{\phi_q} = \frac{(1+q^2)[n_k]+1}{2n_k+1} p_\phi = \Delta_q(n_k) p_\phi \,. \tag{28}$$

Also the commutation relations and plane-wave expansion of the $q$-deformed scalar field $\phi_q(x)$ are given by using (18)-(20) in (7), as follows

$$\phi_q(x)\phi_q^+(x') - q^2\phi_q^+(x')\phi_q(x) = i\Delta(x-x') \,, \tag{29}$$

where

$$\Delta(x-x') = \frac{-1}{(2\pi)^3}\int \frac{d^3k}{w_k} \sin w_k (x - x_0) \,. \tag{30}$$

On the other hand, the deformed and standard annihilation operators, $a_q$ and $a_s$ is written as [62]

$$a_q = a_s \sqrt{\frac{[N_k]}{N_k}} \,, \tag{31}$$

From which we can express the deformed bosonic scalar fields in terms of the standard one by using (21) in (31) and (7):

$$\phi_q = \sqrt{\frac{1-q^{2n_k}}{(1-q^2)n_k}}\,\phi = \Delta(q)\phi \,, \tag{32}$$

where we use the Hermiticity of the number operator $N$ .

Now we will derive the Friedmann equations for our coupling $q$-deformed dark energy to dark matter model with a radiation field in a FRW space-time by using the scale factor $a(t)$ in Einstein's equations. How we can achieve this is to relate the scale factor to the energy-momentum tensor of the objects in the universe under consideration. It is a common fact that to consider energy and matter as a perfect fluid, which will naturally be generalized to dark energy and matter. An isotropic fluid in one coordinate frame gives an isotropic metric in another frame which coincides with the first frame. This means that the fluid is at rest in commoving coordinates. Then the four velocity is given as [63]

$$U^{\mu} = (1,0,0,0)\,, \tag{33}$$

while the energy-momentum tensor reads

$$T_{\mu\nu} = (\rho + p)U_{\mu}U_{\nu} + p\,g_{\mu\nu} = \begin{pmatrix} \rho & 0 & 0 & 0 \\ 0 & & & \\ 0 & & g_{ij}\,p & \\ 0 & & & \end{pmatrix}. \tag{34}$$

Raising one index gives a more suitable form

$$T^{\mu}_{\nu} = diag(-\rho, p, p, p)\,. \tag{35}$$

For a model of universe described by Dirac-Born-Infeld type action and consisting of more than one form of a energy-momentum, we have totally three types of energy density and pressure, such that

$$\rho_{tot} = \rho_{\phi_q} + \rho_m + \rho_r\,, \tag{36}$$

$$p_{tot} = p_{\phi_q} + p_r\,, \tag{37}$$

where the pressure of the dark matter $p_m$ is explicitly zero in the total pressure $p_{tot}$ (37). From the conservation of equation for the zero component $\nabla_{\mu}T^{\mu}_{0} = 0$, one obtains $\rho \propto a^{-3(1+w)}$. Here $w$ is the parameter of the equation of state $p = w\rho$ which relates the pressure and the energy density of the cosmological fluid component under consideration. Therefore, for a matter component pressure is zero, then $w_m = 0$, but for a radiation component due to the vanishing trace of the energy-momentum tensor of the electromagnetic field $w_r = 1/3$. We then express the total equation of state parameter as

$$w_{tot} = \frac{p_{tot}}{\rho_{tot}} = w_{\phi_q}\,\Omega_{\phi_q} + w_r\,\Omega_r\,. \tag{38}$$

While the equation of state parameters are given as $w_{\phi_q} = p_{\phi_q} / \rho_{\phi_q}$, $w_r = p_r / \rho_r = 1/3$, the density parameters are defined by $\Omega_{\phi_q} = \rho_{\phi_q} / \rho_{tot}$, $\Omega_r = \rho_r / \rho_{tot}$ for the *q*-deformed dark energy and the radiation fields, respectively. Since the pressure of the dark matter is $p_m = 0$, then the equation of state parameter is $w_m = p_m / \rho_m = 0$ and the density parameter is $\Omega_m = \rho_m / \rho_{tot}$ for the dark matter field having no contribution to $w_{tot}$ (38), but contributing to total the density parameter, such as

$$\Omega_{tot} = \Omega_{\phi_q} + \Omega_m + \Omega_r = \frac{\kappa^2 \rho_{tot}}{3H^2} = 1\,. \tag{39}$$

We now turn to Einstein's equations of the form $R_{\mu\nu} = \kappa^2 (T_{\mu\nu} - \tfrac{1}{2} g_{\mu\nu} T)$. Then by using the components of the Ricci tensor for a FRW space-time (6) and the energy momentum tensor in (35), we rewrite the Einstein's equations, for $\mu\nu = 00$ and $\mu\nu = i\,j$

$$-3\frac{\ddot{a}}{a} = \frac{\kappa^2}{2}(\rho + 3p)\,, \tag{40}$$

$$\frac{\ddot{a}}{a} + 2\left(\frac{\dot{a}}{a}\right)^2 = \frac{\kappa^2}{2}(\rho - p)\,, \tag{41}$$

respectively. Here dot also represents the derivative with respect to cosmic time *t*. Using (40) and (41) gives the Friedmann equations for FRW metric as

$$H^2 = \frac{\kappa^2}{3}(\rho_{\phi_q} + \rho_m + \rho_r)\,, \tag{42}$$

$$\dot{H} = -\frac{\kappa^2}{2}(\rho_{\phi_q} + p_{\phi_q} + \rho_m + \rho_r + p_r)\,, \tag{43}$$

where $H = \dot{a}/a$ is the Hubble parameter and $\rho_r = 3p_r$. From the conservation of energy, we can obtain the continuity equations for *q*-deformed dark energy, dark matter and the radiation constituents in the form of evolution equations, such as

$$\dot{\rho}_{\phi_q} + 3H(\rho_{\phi_q} + p_{\phi_q}) = -Q'\,, \tag{44}$$

$$\dot{\rho}_m + 3H\rho_m = Q\,, \tag{45}$$

$$\dot{\rho}_r + 3H(\rho_r + p_r) = Q' - Q\,, \tag{46}$$

where $Q$ is an interaction current between the *q*-deformed dark energy and the dark matter which transfers the energy and momentum from the dark matter to dark energy and vice versa. $Q$ and $Q'$ vanish for the models having no coupling between the dark energy and the dark matter. For the models including only the interactions between dark energy and dark matter, the interaction terms become equal $Q' = Q$. The case $Q < 0$ corresponds to an energy transfer from dark matter to the other two constituents, the case $Q' > 0$ corresponds to an energy transfer form dark energy to the other constituents, and the case $Q' < 0$ corresponds to an energy loss from radiation. Here we consider the model only dark energy and dark matter interacts and $Q' = Q$ [64].

The energy density $\rho$ and pressure $p$ of these dark energy is rewritten explicitly from the energy-momentum tensor components (4) and (5) obtained by the Dirac-Born-Infeld type action of the coupling *q*-deformed dark energy and dark matter, such as [65-68]

$$\rho_{\phi_q} = T_0^{0\phi_q} = \frac{1}{2}\dot{\phi}_q^{\,2} + \frac{1}{2}m^2\phi_q^{\,2}\,, \tag{47}$$

$$p_{\phi_q} = -T_i^{i\,\phi_q} = \frac{1}{2}\dot{\phi}_q^{\,2} - \frac{1}{2}m^2\phi_q^{\,2}\,, \tag{48}$$

where the dark energy is space-independent due to the isotropy and homogeneity. Now equation of motion for the $q$-deformed dark energy can be obtained by inserting (47) and (48) into the evolution equation, such as

$$\ddot{\phi}_q + 3H\dot{\phi}_q + m^2\phi_q = -\frac{Q}{\dot{\phi}_q}\,. \tag{49}$$

In order to obtain the energy density, pressure and equation of motion in terms of the deformation parameter $q$, equation (32) and its time derivative will be used. Because the number of particles in each mode of the $q$-deformed scalar field varies in time by the particle creation and annihilation, the time derivative of $\Delta(q)$ is given as

$$\dot{\Delta}(q) = \frac{-q^{2n_k}\dot{n}_k \ln q}{\sqrt{(1-q^2)(1-q^{2n_k})n_k}} - \frac{\dot{n}_k\sqrt{1-q^{2n_k}}}{2\sqrt{(1-q^2)n_k^3}}\,. \tag{50}$$

Substituting (32) and (50) in the equations (47)-(49), we obtain

$$\rho_{\phi_q} = \frac{1}{2}\Delta^2(q)\dot{\phi}^2 + \frac{1}{2}\Delta^2(q)m^2\phi^2 + \frac{1}{2}\dot{\Delta}^2(q)\phi^2 + \Delta(q)\dot{\Delta}(q)\phi\dot{\phi}\,, \tag{51}$$

$$p_{\phi_q} = \frac{1}{2}\Delta^2(q)\dot{\phi}^2 - \frac{1}{2}\Delta^2(q)m^2\phi^2 + \frac{1}{2}\dot{\Delta}^2(q)\phi^2 + \Delta(q)\dot{\Delta}(q)\phi\dot{\phi}\,, \tag{52}$$

$$\Delta(q)\ddot{\phi} + 3\Delta(q)H\dot{\phi} - \Delta(q)m^2\phi + 2\dot{\Delta}(q)\dot{\phi} + \ddot{\Delta}(q)\phi + 3H\dot{\Delta}(q)\phi = -\beta\kappa\rho_m\,. \tag{53}$$

Here, we consider the commonly used interaction current as $Q = \beta\kappa\rho_m\dot{\phi}_q$ in the literature [15], in order to obtain stationary and stable cosmological solutions in our dark model. The deformed energy density and pressure equations (25) and (28) are same with the equations (51) and (52), respectively. While the equations (25) and (51) are the expression of the deformed energy density, accordingly (28) and (52) are the deformed pressure of the dark energy in terms of the deformation parameter *q*. The functions of the deformation parameter in (25) and (51) are

$$\Delta_q(n_k) \approx \Delta(q),$$

$$\frac{(1+q^2)[n_k]+1}{2n_k+1} \approx \sqrt{\frac{1-q^{2n_k}}{(1-q^2)n_k}}, \tag{54}$$

since $n_k$ values are very large, and it is given as a function of time:

$$n_k \approx \frac{t}{3} + \frac{\sqrt{16t^2+16t-14}}{12} + \frac{1}{6}. \tag{55}$$

We now perform the phase-space analysis of our coupling *q*-deformed dark energy to dark matter model if the late-time solutions of the universe can be obtained, in order to confirm our proposal.

### 3. Phase-space analysis

We investigate the cosmological properties of the proposed *q*-deformed dark energy model by performing the phase-space analysis. We need to transform the equations of the dynamical system into its autonomous form [26-28, 36, 37, 69-71]. The auxiliary variables are defined to be

$$x_{\phi_q} = \frac{\kappa(\Delta\dot{\phi} + \dot{\Delta}\phi)}{\sqrt{6}H}, \qquad y_{\phi_q} = \frac{\kappa\sqrt{e^{-\kappa\lambda\Delta\phi}}}{\sqrt{3}H}, \tag{56}$$

We consider an exponential potential as $V = V_0 e^{-\kappa\lambda\phi_q}$ instead of the potential $V = (1/2)m^2\phi_q{}^2$ in the Lagrangian (1), as the usual assumption in the literature, because the power-law potential does not provide a stable attractor solution [16, 18, 72-76].

We also express the density parameters for the *q*-deformed scalar field dark energy, dark matter and the radiation in the autonomous system by using (36), (37) and (51) with (56)

$$\Omega_{\phi_q} = \frac{\kappa^2 \rho_{\phi_q}}{3H^2} = x_{\phi_q}^2 + y_{\phi_q}^2 , \tag{57}$$

$$\Omega_m = \frac{\kappa^2 \rho_m}{3H^2} , \tag{58}$$

$$\Omega_r = \frac{\kappa^2 \rho_r}{3H^2} , \tag{59}$$

then the total density parameter is given by

$$\Omega_{tot} = \frac{\kappa^2 \rho_{tot}}{3H^2} = x_{\phi_q}^2 + y_{\phi_q}^2 + \Omega_m + \Omega_r = 1, \tag{60}$$

Also, the equation of state parameter for the dark energy is written in the autonomous form by using (36) and (37) with (56)

$$w_{\phi_q} = \frac{p_{\phi_q}}{\rho_{\phi_q}} = \frac{x_{\phi_q}^2 - y_{\phi_q}^2}{x_{\phi_q}^2 + y_{\phi_q}^2} . \tag{61}$$

Then the total equation of state parameter in the autonomous system from (38) and (57)-(59) and (61) is obtained as

$$w_{tot} = x_{\phi_q}^2 - y_{\phi_q}^2 + \frac{\Omega_r}{3} . \tag{62}$$

We also define $s = -\dot{H}/H^2$ in the autonomous system by using (42), (43) and (62), such that

$$s = -\frac{\dot{H}}{H^2} = \frac{3}{2}(1 + w_{tot}) = \frac{3}{2}\left(1 + x_{\phi_q}^2 - y_{\phi_q}^2 + \frac{\Omega_r}{3}\right). \tag{63}$$

$s$ is here only a jerk parameter which is used in other equations of cosmological parameters. However the deceleration parameter $q_d$ which is not used in the equations, but not also a jerk parameter is defined as

$$q_d = -1 - \frac{\dot{H}}{H^2}. \tag{64}$$

Now we convert the Friedmann equations (42), (43), the continuity equations (45) and (46), and the equation of motion (53) into the autonomous system by using the auxiliary variables in (56)-(59) and their derivatives with respect to $N = \ln a$, for which the time derivative of any quantity $F$ is $\dot{F} = H(dF/dN)$. Thus we will obtain $X' = f(X)$, where $X$ is the column vector including the auxiliary variables and $f(X)$ is the column vector of the autonomous equations. We then find the critical points $X_c$ of $X$, by setting $X' = 0$. We then expand $X' = f(X)$ around $X = X_c + U$, where $U$ is the column vector of perturbations of the auxiliary variables, such as $\delta x_{\phi_q}$, $\delta y_{\phi_q}$, $\delta\Omega_m$ and $\delta\Omega_r$ for each constituents in our model. Thus, we expand the perturbation equations up to the first order for each critical point as $U' = MU$, where $M$ is the matrix of perturbation equations. The eigenvalues of perturbation matrix $M$ determine the type and stability of the critical points for each critical points [77-79].

With the definitions for the interaction current and the potential, the autonomous form of the cosmological system reads [80-89]

$$x'_{\phi_q} = -3x_{\phi_q} + s x_{\phi_q} + \frac{\sqrt{6}}{2}\lambda y_{\phi_q}^2 - \frac{\sqrt{6}}{2}\beta\Omega_m, \tag{65}$$

$$y'_{\phi_q} = s\, y_{\phi_q} - \frac{\sqrt{6}}{2}\lambda y_{\phi_q} x_{\phi_q}, \tag{66}$$

$$\Omega_m' = \Omega_m[-3+\sqrt{6}\beta x_{\phi_q} + 2s], \tag{67}$$

$$\Omega_r' = \Omega_r[-4+2s]. \tag{68}$$

In order to perform the phase-space analysis of the model, we obtain the critical points of the autonomous system in (65)-(68). We will obtain these points by equating the left hand sides of the equations (65)-(68) to zero for stationary solutions, by using the conditions $\Omega_{tot} = 1$. After some calculations, five critical points are found as listed in Table 1 with the existence conditions.

Now we will get the perturbations $\delta x_{\phi_q}'$, $\delta y_{\phi_q}'$, $\delta \Omega_m'$ and $\delta \Omega_r'$ for each constituents in our model by using the variations of the equations (65)-(68), such as

$$\delta x_{\phi_q}' = \left[-\frac{3}{2}+\frac{9}{2}x_{\phi_q}^2 - \frac{3}{2}y_{\phi_q}^2 + \frac{\Omega_r}{2}\right]\delta x_{\phi_q} + \left[\sqrt{6}\lambda - 3x_{\phi_q} y_{\phi_q}\right]\delta y_{\phi_q} + \frac{\sqrt{6}}{2}\beta\,\delta\Omega_m + \frac{x_{\phi_q}}{2}\delta\Omega_r, \tag{69}$$

$$\delta y_{\phi_q}' = \left[3x_{\phi_q} y_{\phi_q} - \frac{\sqrt{6}}{2}\lambda y_{\phi_q}\right]\delta x_{\phi_q} + \left[\frac{3}{2}+\frac{\sqrt{6}}{2}\lambda x_{\phi_q} + \frac{3}{2}x_{\phi_q}^2 - \frac{9}{2}y_{\phi_q}^2 + \frac{\Omega_r}{2}\right]\delta y_{\phi_q} + \frac{y_{\phi_q}}{2}\delta\Omega_r, \tag{70}$$

$$\begin{aligned}\delta\Omega_m' = {} & [6x_{\phi_q}\Omega_m + \sqrt{6}\beta\Omega_m]\delta x_{\phi_q} - 6y_{\phi_q}\Omega_m\,\delta y_{\phi_q} + [\sqrt{6}\beta x_{\phi_q} + 3x_{\phi_q}^2 - 3y_{\phi_q}^2 + \Omega_r]\delta\Omega_m \\ & + \Omega_m\,\delta\Omega_r\end{aligned}, \tag{71}$$

$$\delta\Omega_r' = 6x_{\phi_q}\Omega_r\delta x_{\phi_q} - 6y_{\phi_q}\Omega_r\,\delta y_{\phi_q} + [-1+3x_{\phi_q}^2 - 3y_{\phi_q}^2 + 2\Omega_r]\delta\Omega_r\ . \tag{72}$$

Thus we obtain a $4\times 4$ perturbation matrix $M$ whose non-zero elements are given as

$$M_{11} = -\frac{3}{2}+\frac{9}{2}x_{\phi_q}^2 - \frac{3}{2}y_{\phi_q}^2 + \frac{\Omega_r}{2},$$

$$M_{12} = \sqrt{6}\lambda y_{\phi_q} - 3x_{\phi_q} y_{\phi_q},$$

$$M_{13} = \frac{\sqrt{6}}{2}\beta,$$

$$M_{14} = \frac{x_{\phi_q}}{2} ,$$

$$M_{21} = 3x_{\phi_q} y_{\phi_q} - \frac{\sqrt{6}}{2} \lambda y_{\phi_q} ,$$

$$M_{22} = \frac{3}{2} + \frac{\sqrt{6}}{2} \lambda x_{\phi_q} + \frac{3}{2} x_{\phi_q}^2 - \frac{9}{2} y_{\phi_q}^2 + \frac{\Omega_r}{2} ,$$

$$M_{24} = \frac{y_{\phi_q}}{2} ,$$

$$M_{31} = 6x_{\phi_q} \Omega_m + \sqrt{6}\beta \Omega_m ,$$

$$M_{32} = -6y_{\phi_q} \Omega_m ,$$

$$M_{33} = \sqrt{6}\beta x_{\phi_q} + 3x_{\phi_q}^2 - 3y_{\phi_q}^2 + \Omega_r ,$$

$$M_{34} = \Omega_m ,$$

$$M_{41} = 6x_{\phi_q} \Omega_r ,$$

$$M_{42} = -6y_{\phi_q} \Omega_r ,$$

$$M_{44} = -1 + 3x_{\phi_q}^2 - 3y_{\phi_q}^2 + 2\Omega_r . \tag{73}$$

Then we insert linear perturbations $x_{\phi_q} \to x_{\phi_q,c} + \delta x_{\phi_q}$, $y_{\phi_q} \to y_{\phi_q,c} + \delta y_{\phi_q}$, $\Omega_m \to \Omega_{m,c} + \delta \Omega_m$ and $\Omega_r \to \Omega_{r,c} + \delta \Omega_r$ about the critical points for three constituents in the autonomous system (65)-(68). So that we can calculate the eigenvalues of perturbation matrix $M$ for five critical points given in Table 1, with the corresponding existing conditions.

We find and represent five perturbation matrices in Table 2 for each of the five critical points. We obtain five sets of eigenvalues. In order to determine the type and stability of critical points, we investigate the sign of the real parts of eigenvalues. A critical point is stable if all the real part of eigenvalues is negative. The physical meaning of the negative eigenvalue is always stable attractor, namely if the universe is in this state, it keeps its state forever and thus it can attract the universe at a late-time. There can occur accelerated expansion only for $w_{tot} < -1/3$.

Eigenvalues of the five $M$ matrices with the existence conditions, stability conditions and acceleration condition are represented in Table 3, for each critical points $A$, $B$, $C$, $D$ and $E$. As seen in Table 3, the first two critical points $A$ and $B$ have the same eigenvalues. Here the eigenvalues and the stability conditions of the perturbation matrices for critical points $A$, $B$, $D$ and $E$ have been obtained by the numerical methods, since the complexity of the matrices. The stability conditions of each critical point are listed in Table 3, according to the sign of the eigenvalues.

We now analyze the cosmological behavior of each critical point by noting the attractor solutions in scalar field cosmology [90]. From the theoretical cosmology studies, we know that the energy density of a scalar field has effect on the determination of the evolution of universe. Cosmological attractors provide the understanding the evolution and the affecting factors on this evolution, such as, from the dynamical conditions the scalar field evolution approaches a certain kind of behavior without initial fine tuning conditions [91–101]. Attractor behavior is known as a situation in which a collection of phase-space points evolve into a particular region and never leave from there.

*Critical point A:* This point exists for $\beta(\lambda+\beta) > -3/2$ and $\lambda(\lambda+\beta) > 3$. Because of $w_{tot} < -1/3$, acceleration occurs at this point if $\lambda < 2\beta$ and it is an expansion phase since $y_{\phi_q}$ is positive, so $H$ is positive, too. Point $A$ is stable meaning that universe keeps its further evolution, if $\lambda$ and $\beta$ take the values for the negative eigenvalues given in the second column of Table3. In Figure 1, we also represent the 2-dimensional and 3-dimensional projections of 4-dimensional phase-space trajectories for $\beta = 2.51$, $\lambda = 3.1$, $\lambda = 3.6$ and $\lambda = 4.1$. This state corresponds to a stable attractor starting from the critical point $A = (0.21, 0.70, 0.45, 0)$, as seen from the plots in Figure 1. Also zero value of critical point $\Omega_r$ cancels the total behavior $\Omega'_r$ in (72).

*Critical point B:* Point $B$ also exists for $\beta(\lambda+\beta) > -3/2$ and $\lambda(\lambda+\beta) > 3$. Acceleration phase is again valid here if $\lambda < 2\beta$ leading $w_{tot} < -1/3$, but this point refers to contraction phase because $y_{\phi_q}$ is negative here. Stability of the point $B$ is again satisfied for $\lambda$ and $\beta$ values given in the second column of Table3. Therefore, it is represented that the stable attractor behavior starting from the critical point $B=(0.21,-0.70,0.45,0)$ for $\beta=2.51$, $\lambda=3.1$, $\lambda=3.6$ and $\lambda=4.1$ values, in Figure 2. The zero value of critical points $\Omega_r$ again cancels the total behavior $\Omega_r'$ in (72).

*Critical point C:* Critical point $C$ occurs for all values of $\beta$, while $\lambda<\sqrt{6}$. The cosmological behavior is again an acceleration phase occurs if $\lambda<\sqrt{2}$ providing $w_{tot}<-1/3$ and an expansion phase since $y_{\phi_q}$ is positive. Point $C$ is stable if $\beta<(3-\lambda^2)/\lambda$ and $\lambda<\sqrt{3}$ . 2-dimensional projection of phase-space are represented in Figure 3, for $\beta=1.5$, $\lambda=0.001$, $\lambda=0.5$ and $\lambda=1.1$. The stable attractor starting from the critical point $C=(0.48,0.87,0,0)$ can be inferred from the Figure 3. We again find zero plots containing zero values $\Omega_m$ and $\Omega_r$, since it cancels the total behavior $\Omega_m'$ and $\Omega_r'$ in (71) and (72).

*Critical point D:* This point exists for any values of $\beta$, while $\lambda>2$. Acceleration phase never occurs due to $w_{tot}=1/3$. Point $D$ is always unstable for any values of $\beta$ and $\lambda$. This state corresponds to a unstable saddle point starting from the point $D=(0.54,0.38,0,0.55)$ for $\beta=1.5$, $\lambda=3$, $\lambda=4$ and $\lambda=6$, as seen from the plots in Figure 4. Zero plots containing the axis $\Omega_m$ leads the cancellation of the total behavior $\Omega_m'$ in (71), since $\Omega_m=0$, so they are not represented in the Figure 4.

*Critical point E:* This point exists for any values of $\lambda$, while $\beta>1/\sqrt{2}$. Acceleration phase never occurs due to $w_{tot}=1/3$. Point $E$ is always unstable for any values of $\beta$ and $\lambda$. This state corresponds to a unstable saddle point starting from the point $E=(-0.24,0,0.11,0.82)$ $\lambda=1$, $\beta=1.7$, $\beta=2.6$ and $\beta=3.5$, as seen from the plots in Figure 5. Zero plots containing the axis $y_{\phi_q}$ leads the cancellation of the total behavior $y_{\phi_q}'$ in (70), since $y_{\phi_q}=0$, so they are not represented in the Figure 5.

All the plots in Figures 1-3 have the structure of stable attractor, since each of them evolves to a single point which is in fact one of the critical points in Table 1. These evolutions to the critical points are the attractor solutions in coupling *q*-deformed dark energy to dark matter cosmology of our model, which imply an expanding universe. Therefore, we confirm that the dark energy in our model can be defined in terms of the *q*-deformed scalar fields obeying the *q*-deformed boson algebra in (18)-(21). According to the stable attractor behaviors, it makes sense to consider the dark energy as a scalar field defined by the *q*-deformed scalar field, since the negative pressure of *q*-deformed boson field, as dark energy field.

Finally we can investigate the relation between *q*-deformed and standard dark energy density, pressure and scalar field equations in (25), (28) and (32). We illustrate the behavior of *q*-deformed energy density and pressure in terms of the standard ones with respect to the total number of particles and the deformation parameter $q$ in Figure 6 and Figure 7, respectively. We observe that for the large particle number the *q*-deformed energy density and pressure function decrease with the decrease in deformation parameter *q*. Contrary, if the particle number is small, the deformed energy density and pressure increase with the decrease in deformation parameter. Note that, when the deformation parameter decreases from 1, this increases the deformation of the model, since the deformation vanishes by approaching 1. The deformation parameter significantly affects the value of the deformed energy density and pressure. In the $q \to 1$ limit, deformed energy density and pressure function becomes identical to the standard values, as expected.

In Figure 8, we represent the *q*-deformed scalar field behavior in terms of the standard one. It is observed that while the deformation parameter $q \to 1$, *q*-deformed scalar field becomes identical to the standard one. However, it asymptotically approaches to lower values, while $q$ decreases with large number of particles. Since the square of a quantum mechanical field means the probability density, *q*-deformed probability density decreases when the deformation increases, and in the $q \to 0$ limit it approaches zero.

Also, since the dark matter pressure is taken to be zero, $\omega_m \approx 0$ and $\omega_{tot} \approx \omega_{\phi_q}$. For the stable accelerated expansion condition $\omega_{tot} \approx \omega_{\phi_q} < -1/3$, solutions requires the scalar field dark energy pressure to be negative $p_{\phi_q} < 0$. From the relation $p_{\phi_q} = \Delta_q(n_k) p_\phi$ in (28), we finally represent the effect of $q$ on the deformed dark energy pressure $p_{\phi_q}$, namely on the accelerated expansion behavior in Figure 9. From the figure, we deduce that for any values of $q$, the deformed dark energy shows the accelerated expansion behavior with the negative deformed dark energy pressure.

**4. Conclusion**

Since it is known that the dark energy has a negative pressure acting as a gravitational repulsion to drive the accelerated expansion of universe, we are motivated to propose that the dark energy consists of negative-pressure *q*-deformed scalar field whose field equation is defined by the *q* annihilation and creation operators satisfying the *q*-deformed boson algebra in (18)-(21). In order to confirm our proposal, we consider a *q*-deformed dark energy coupling to the dark matter inhomogeneities, and then investigate the dynamics of the model and phase-space analysis whether it will give stable attractor solutions meaning indirectly an accelerating expansion phase of universe. Therefore, the action integral of coupling *q*-deformed dark energy model is set up to study its dynamics, and the Hubble parameter and Friedmann equations of model are obtained in spatially-flat FRW geometry. Later on, we find the energy density and pressure values with the evolution equations for *q*-deformed dark energy, dark matter and the radiation fields from the variation of the action and the Lagrangian of model. After that we translate these dynamical equations into the autonomous form by introducing suitable auxiliary variables, in order to perform the phase-space analysis of the model. Then the critical points of autonomous system are obtained by setting each autonomous equation to zero and four perturbation matrices can be written for each critical point by constructing the perturbation equations. We determine the eigenvalues of four perturbation matrices to examine the stability of critical points. There are also calculated some important cosmological parameters, such as the total equation of state parameter and the deceleration parameter to check whether the critical points satisfy an accelerating universe. We obtain four stable attractors for the model depending on the coupling parameter $\beta$. An accelerating universe exists for all stable solutions due to $w_{tot} < -1/3$. The critical points $A$ and $B$ are late-time stable attractors for the given $\lambda$ and $\beta$ values for the negative eigenvalues in the second column of Table 3. The point $A$ refers to an expansion, while the point $B$ refers to a contraction with a stable acceleration for $\lambda < 2\beta$. However, the critical points $C$ is late-time stable attractors for $\beta < (3-\lambda^2)/\lambda$ and $\lambda < \sqrt{3}$ with an expansion. The stable attractor behavior of the model at each critical point is demonstrated in Figures 1-3. In order to solve the differential equation system (65)-(68) with the critical points and plot the graphs in Figures 1-5, we use adaptive Runge-Kutta method of 4th and 5th order, in Matlab programming. Then the solutions with the stability conditions of critical points are plotted for each pair of the solution set being the auxiliary variables in (56), (58) and (59).

These figures represent that by choosing the suitable parameters of the model, we obtain the stable and unstable attractors as $A$, $B$, $C$, $D$ and $E$, depending on the existence conditions of critical points $A$, $B$, $C$ $D$ and $E$. Also the suitable parameters with the stability conditions give the stable accelerated behavior for $A$, $B$ and $C$ attractor models.

The $q$-deformed dark energy is a generalization of the standard scalar field dark energy. As seen from the behavior of the deformed energy density, pressure and scalar field functions with respect to the standard functions, in the $q \to 1$ limit they all approach to the standard corresponding function values. However, in the $q \to 0$ limit the deformed energy density and the pressure functions decreases to smaller values of the standard energy density and the pressure function values, respectively. This implies that the energy-momentum of the scalar field decreases when the deformation becomes more apparent, since $q$ reaches 1 gives the non-deformed state. Also when $q \to 0$ for large $n$ values, the deformed scalar field approaches to zero value meaning a decrease in the probability density of the scalar field. This state is expected to represent an energy-momentum decrease leading to a decrease in the probability of the finding the particles of the field. Consequently, $q$ deformation of the scalar field dark energy gives a self-consistent model due to the existence of standard case parameters of the dark energy in the $q \to 1$ limit and the existence of the stable attractor behavior of the accelerated expansion phase of universe for the considered coupling dark energy and dark matter model.

The results confirm that the proposed $q$-deformed scalar field dark energy model is consistent since it gives the expected behavior of the universe. The idea to consider the dark energy as a $q$-deformed scalar field is a very recent approach. There are more deformed particle algebras in the literature which can be considered as other and maybe more suitable candidates for the dark energy. As a further study on the purpose of confirmation whether the dark energy can be considered as a general deformed scalar field, it can be investigated that the other couplings between dark energy and dark matter, and also in the other framework of gravity, such as teleparallel or maybe f(R) gravity.

TABLE 1: Critical points and existence conditions

| Label | $x_{\phi_q}$ | $y_{\phi_q}$ | $\Omega_m$ | $\Omega_r$ | $\Omega_{\phi_q}$ | $\omega_{\phi_q}$ | $\omega_{tot}$ |
|---|---|---|---|---|---|---|---|
| $A$ | $\frac{3}{\sqrt{6}(\lambda+\beta)}$ | $\frac{\sqrt{2\beta(\lambda+\beta)+3}}{\sqrt{2}(\lambda+\beta)}$ | $\frac{\lambda(\lambda+\beta)-3}{(\lambda+\beta)^2}$ | 0 | $\frac{\beta(\lambda+\beta)+3}{(\lambda+\beta)^2}$ | $\frac{-\beta(\lambda+\beta)}{\beta(\lambda+\beta)+3}$ | $\frac{-\beta}{(\lambda+\beta)}$ |
| $B$ | $\frac{3}{\sqrt{6}(\lambda+\beta)}$ | $\frac{-\sqrt{2\beta(\lambda+\beta)+3}}{\sqrt{2}(\lambda+\beta)}$ | $\frac{\lambda(\lambda+\beta)-3}{(\lambda+\beta)^2}$ | 0 | $\frac{\beta(\lambda+\beta)+3}{(\lambda+\beta)^2}$ | $\frac{-\beta(\lambda+\beta)}{\beta(\lambda+\beta)+3}$ | $\frac{-\beta}{(\lambda+\beta)}$ |
| $C$ | $\frac{\sqrt{6}\lambda}{6}$ | $\sqrt{1-\frac{\lambda^2}{6}}$ | 0 | 0 | 1 | $\frac{\lambda^2}{3}-1$ | $\frac{\lambda^2}{3}-1$ |
| $D$ | $\frac{4}{\sqrt{6}\lambda}$ | $\frac{2}{\sqrt{3}\lambda}$ | 0 | $1-\frac{4}{\lambda^2}$ | $\frac{4}{\lambda^2}$ | $\frac{1}{3}$ | $\frac{1}{3}$ |
| $E$ | $\frac{-1}{\sqrt{6}\beta}$ | 0 | $\frac{1}{3\beta^2}$ | $1-\frac{1}{2\beta^2}$ | $\frac{1}{6\beta^2}$ | 1 | $\frac{1}{3}$ |

TABLE 2: Perturbation matrices for each critical point

| Critical Points | Perturbation Matrices |
|---|---|
| $A$ | $M=\begin{pmatrix} \frac{9}{2(\lambda+\beta)^2}+\frac{3\lambda}{2(\lambda+\beta)}-3 & \left[\frac{-9}{6(\lambda+\beta)}+\sqrt{6}\lambda\right]\frac{\sqrt{2\beta(\lambda+\beta)+3}}{\sqrt{2}(\lambda+\beta)} & \frac{-\sqrt{6}\beta}{2} & \frac{\sqrt{6}}{4(\lambda+\beta)} \\ \left[\frac{9}{6(\lambda+\beta)}-\frac{\sqrt{6}\lambda}{2}\right]\frac{\sqrt{2\beta(\lambda+\beta)+3}}{\sqrt{2}(\lambda+\beta)} & \frac{-3(2\beta(\lambda+\beta)+3)}{2(\lambda+\beta)^2} & 0 & \frac{\sqrt{2\beta(\lambda+\beta)+3}}{2\sqrt{2}(\lambda+\beta)} \\ \left[\frac{18}{\sqrt{6}(\lambda+\beta)}+\sqrt{6}\beta\right]\frac{\lambda(\lambda+\beta)-3}{(\lambda+\beta)^2} & \frac{-6(\lambda(\lambda+\beta)-3)\sqrt{2\beta(\lambda+\beta)+3}}{\sqrt{2}(\lambda+\beta)^3} & 0 & \frac{\lambda(\lambda+\beta)-3}{(\lambda+\beta)^2} \\ 0 & 0 & 0 & -4+\frac{3\lambda}{\lambda+\beta} \end{pmatrix}$ |
| $B$ | $M=\begin{pmatrix} \frac{9}{2(\lambda+\beta)^2}+\frac{3\lambda}{2(\lambda+\beta)}-3 & \left[\frac{9}{6(\lambda+\beta)}-\sqrt{6}\lambda\right]\frac{\sqrt{2\beta(\lambda+\beta)+3}}{\sqrt{2}(\lambda+\beta)} & \frac{-\sqrt{6}\beta}{2} & \frac{\sqrt{6}}{4(\lambda+\beta)} \\ \left[\frac{-9}{6(\lambda+\beta)}+\frac{\sqrt{6}\lambda}{2}\right]\frac{\sqrt{2\beta(\lambda+\beta)+3}}{\sqrt{2}(\lambda+\beta)} & \frac{-3(2\beta(\lambda+\beta)+3)}{2(\lambda+\beta)^2} & 0 & \frac{-\sqrt{2\beta(\lambda+\beta)+3}}{2\sqrt{2}(\lambda+\beta)} \\ \left[\frac{18}{\sqrt{6}(\lambda+\beta)}+\sqrt{6}\beta\right]\frac{\lambda(\lambda+\beta)-3}{(\lambda+\beta)^2} & \frac{6(\lambda(\lambda+\beta)-3)\sqrt{2\beta(\lambda+\beta)+3}}{\sqrt{2}(\lambda+\beta)^3} & 0 & \frac{\lambda(\lambda+\beta)-3}{(\lambda+\beta)^2} \\ 0 & 0 & 0 & -4+\frac{3\lambda}{\lambda+\beta} \end{pmatrix}$ |
| $C$ | $M=\begin{pmatrix} \lambda^2-3 & \frac{\sqrt{6}\lambda}{2}\sqrt{1-\frac{\lambda^2}{6}} & \frac{-\sqrt{6}\beta}{2} & \frac{\sqrt{6}\lambda}{12} \\ 0 & \frac{\lambda^2}{2}-3 & 0 & \frac{1}{2}\sqrt{1-\frac{\lambda^2}{6}} \\ 0 & 0 & \lambda^2+\lambda\beta-3 & 0 \\ 0 & 0 & 0 & \lambda^2-4 \end{pmatrix}$ |
| $D$ | $M=\begin{pmatrix} \frac{8}{\lambda^2}-1 & 2\sqrt{2}-\frac{4\sqrt{2}}{\lambda^2} & \frac{-\sqrt{6}\beta}{2} & \frac{2}{\sqrt{6}\lambda} \\ -\sqrt{2}+\frac{4\sqrt{2}}{\lambda^2} & \frac{4}{\lambda^2} & 0 & \frac{1}{\sqrt{3}\lambda} \\ 0 & 0 & \frac{4\beta}{\lambda}+1 & 0 \\ \frac{24}{\sqrt{6}\lambda}(1-\frac{4}{\lambda^2}) & \frac{-12}{\sqrt{3}\lambda}(1-\frac{4}{\lambda^2}) & 0 & 1-\frac{4}{\lambda^2} \end{pmatrix}$ |
| $E$ | $M=\begin{pmatrix} \frac{1}{2\beta^2}-1 & 0 & \frac{-\sqrt{6}\beta}{2} & \frac{-1}{2\sqrt{6}\beta} \\ 0 & 2+\frac{\lambda}{\beta} & 0 & 0 \\ 0 & 0 & 0 & \frac{1}{3\beta^2} \\ \frac{\sqrt{6}}{\beta}(1-\frac{1}{2\beta^2}) & 0 & 0 & 1-\frac{1}{2\beta^2} \end{pmatrix}$ |

TABLE 3: Eigenvalues and stability of critical points

<table>
<tr><td rowspan="25">A and B</td><td colspan="4">Eigenvalues</td><td>$\lambda$</td><td>$\beta$</td><td rowspan="25">Existing condition is $\beta(\lambda+\beta) > -3/2$ and $\lambda(\lambda+\beta) > 3$.<br><br>Stable point if $\lambda$ and $\beta$ are the given values for the negative eigenvalues in the second column.<br><br>Acceleration phase occurs if $\lambda < 2\beta$ .</td></tr>
<tr><td>-0.7516</td><td>-0.7516</td><td>-0.0065</td><td>-1.0065</td><td>4.6000</td><td>0.0100</td></tr>
<tr><td>-0.7518</td><td>-0.7518</td><td>-0.0073</td><td>-1.0073</td><td>4.1000</td><td>0.0100</td></tr>
<tr><td>-0.7521</td><td>-0.7521</td><td>-0.0083</td><td>-1.0083</td><td>3.6000</td><td>0.0100</td></tr>
<tr><td>-0.7524</td><td>-0.7524</td><td>-0.0096</td><td>-1.0096</td><td>3.1000</td><td>0.0100</td></tr>
<tr><td>-0.8249</td><td>-0.8249</td><td>-0.2994</td><td>-1.2994</td><td>4.6000</td><td>0.5100</td></tr>
<tr><td>-0.8330</td><td>-0.8330</td><td>-0.3319</td><td>-1.3319</td><td>4.1000</td><td>0.5100</td></tr>
<tr><td>-0.8431</td><td>-0.8431</td><td>-0.3723</td><td>-1.3723</td><td>3.6000</td><td>0.5100</td></tr>
<tr><td>-0.8560</td><td>-0.8560</td><td>-0.4238</td><td>-1.4238</td><td>3.1000</td><td>0.5100</td></tr>
<tr><td>-0.8850</td><td>-0.8850</td><td>-0.5401</td><td>-1.5401</td><td>4.6000</td><td>1.0100</td></tr>
<tr><td>-0.8982</td><td>-0.8982</td><td>-0.5930</td><td>-1.5930</td><td>4.1000</td><td>1.0100</td></tr>
<tr><td>-0.9143</td><td>-0.9143</td><td>-0.6573</td><td>-1.6573</td><td>3.6000</td><td>1.0100</td></tr>
<tr><td>-0.9343</td><td>-0.9343</td><td>-0.7372</td><td>-1.7372</td><td>3.1000</td><td>1.0100</td></tr>
<tr><td>-0.9354</td><td>-0.9354</td><td>-0.7414</td><td>-1.7414</td><td>4.6000</td><td>1.5100</td></tr>
<tr><td>-0.9519</td><td>-0.9519</td><td>-0.8075</td><td>-1.8075</td><td>4.1000</td><td>1.5100</td></tr>
<tr><td>-0.9716</td><td>-0.9716</td><td>-0.8865</td><td>-1.8865</td><td>3.6000</td><td>1.5100</td></tr>
<tr><td>-0.9781</td><td>-0.9781</td><td>-0.9123</td><td>-1.9123</td><td>4.6000</td><td>2.0100</td></tr>
<tr><td>-0.9957</td><td>-0.9957</td><td>-0.9826</td><td>-1.9826</td><td>3.1000</td><td>1.5100</td></tr>
<tr><td>-0.9967</td><td>-0.9967</td><td>-0.9869</td><td>-1.9869</td><td>4.1000</td><td>2.0100</td></tr>
<tr><td>-1.0148</td><td>-1.0148</td><td>-1.0591</td><td>-2.0591</td><td>4.6000</td><td>2.5100</td></tr>
<tr><td>-1.0187</td><td>-1.0187</td><td>-1.0749</td><td>-2.0749</td><td>3.6000</td><td>2.0100</td></tr>
<tr><td>-1.0348</td><td>-1.0348</td><td>-1.1392</td><td>-2.1392</td><td>4.1000</td><td>2.5100</td></tr>
<tr><td>-1.0450</td><td>-1.0450</td><td>-1.1800</td><td>-2.1800</td><td>3.1000</td><td>2.0100</td></tr>
<tr><td>-1.0581</td><td>-1.0581</td><td>-1.2324</td><td>-2.2324</td><td>3.6000</td><td>2.5100</td></tr>
<tr><td>-1.0856</td><td>-1.0856</td><td>-1.3422</td><td>-2.3422</td><td>3.1000</td><td>2.5100</td></tr>
<tr><td>C</td><td colspan="6">Eigenvalues<br>$\lambda^2+\lambda\beta-3$,<br>$\lambda^2-3$,<br>$\lambda^2-4$,<br>$\frac{\lambda^2}{2}-3$,</td><td>Existing condition is $\lambda < \sqrt{6}$ .<br><br>Stable point if $\beta < \frac{3-\lambda^2}{\lambda}$ and $\lambda < \sqrt{3}$<br><br>Acceleration phase occurs if $\lambda < \sqrt{2}$ .</td></tr>
</table>

| | Eigenvalues | | | | $\lambda$ | $\beta$ | |
|---|---|---|---|---|---|---|---|
| | -0.5000 | -0.5000 | 1.0000 | 0.2000 | 2.5000 | -0.5000 | |
| | -0.5000 | -0.5000 | 1.0000 | 0.3333 | 3.0000 | -0.5000 | |
| | -0.5000 | -0.5000 | 1.0000 | 0.4286 | 3.5000 | -0.5000 | |
| | -0.5000 | -0.5000 | 1.0000 | 0.5000 | 4.0000 | -0.5000 | |
| | 1.0000 | -1.0000 | 0 | 0 | 2.0000 | -0.5000 | |
| | 1.0000 | -1.0000 | 2.0000 | 0 | 2.0000 | 0.5000 | |
| | 1.0000 | -1.0000 | 4.0000 | 0 | 2.0000 | 1.5000 | |
| | 1.0000 | -1.0000 | -2.0000 | 0 | 2.0000 | -1.5000 | |
| | -0.5000 | -0.5000 | 1.0000 | 0 | 4.0000 | -1.0000 | |
| | -0.5000 | -0.5000 | 1.0000 | 1.0000 | 2.5000 | 0 | |
| | -0.5000 | -0.5000 | 1.0000 | 1.0000 | 3.5000 | 0 | |
| | -0.5000 | -0.5000 | 1.0000 | 1.0000 | 4.0000 | 0 | |
| | -0.5000 | -0.5000 | 1.0000 | 1.5000 | 4.0000 | 0.5000 | |
| | -0.5000 | -0.5000 | 1.0000 | 1.5714 | 3.5000 | 0.5000 | |
| | -0.5000 | -0.5000 | 1.0000 | 1.6667 | 3.0000 | 0.5000 | Existing condition is $\lambda > 2$ and |
| | -0.5000 | -0.5000 | 1.0000 | 1.8000 | 2.5000 | 0.5000 | $\forall \beta$ . |
| | -0.5000 | -0.5000 | 1.0000 | 2.0000 | 4.0000 | 1.0000 | |
| | -0.5000 | -0.5000 | 1.0000 | 2.1429 | 3.5000 | 1.0000 | |
| *D* | -0.5000 | -0.5000 | 1.0000 | 2.3333 | 3.0000 | 1.0000 | Unstable point. |
| | -0.5000 | -0.5000 | 1.0000 | 2.5000 | 4.0000 | 1.5000 | |
| | -0.5000 | -0.5000 | 1.0000 | 2.6000 | 2.5000 | 1.0000 | |
| | -0.5000 | -0.5000 | 1.0000 | 2.7143 | 3.5000 | 1.5000 | Acceleration phase never |
| | -0.5000 | -0.5000 | 1.0000 | 3.0000 | 3.0000 | 1.5000 | occurs. |
| | -0.5000 | -0.5000 | 1.0000 | 3.0000 | 4.0000 | 2.0000 | |
| | -0.5000 | -0.5000 | 1.0000 | 3.2857 | 3.5000 | 2.0000 | |
| | -0.5000 | -0.5000 | 1.0000 | 3.4000 | 2.5000 | 1.5000 | |
| | -0.5000 | -0.5000 | 1.0000 | 3.6667 | 3.0000 | 2.0000 | |
| | -0.5000 | -0.5000 | 1.0000 | 4.2000 | 2.5000 | 2.0000 | |
| | -0.5000 | -0.5000 | 1.0000 | -0.1429 | 3.5000 | -1.0000 | |
| | -0.5000 | -0.5000 | 1.0000 | -0.3333 | 3.0000 | -1.0000 | |
| | -0.5000 | -0.5000 | 1.0000 | -0.5000 | 4.0000 | -1.5000 | |
| | -0.5000 | -0.5000 | 1.0000 | -0.6000 | 2.5000 | -1.0000 | |
| | -0.5000 | -0.5000 | 1.0000 | -0.7143 | 3.5000 | -1.5000 | |
| | -0.5000 | -0.5000 | 1.0000 | -1.0000 | 3.0000 | -1.5000 | |
| | -0.5000 | -0.5000 | 1.0000 | -1.0000 | 4.0000 | -2.0000 | |
| | -0.5000 | -0.5000 | 1.0000 | -1.2857 | 3.5000 | -2.0000 | |
| | -0.5000 | -0.5000 | 1.0000 | -1.4000 | 2.5000 | -1.5000 | |
| | -0.5000 | -0.5000 | 1.0000 | -1.6667 | 3.0000 | -2.0000 | |
| | -0.5000 | -0.5000 | 1.0000 | -2.2000 | 2.5000 | -2.0000 | |

| | Eigenvalues | | | | $\lambda$ | $\beta$ | |
|---|---|---|---|---|---|---|---|
| | 0.2634 | -0.1317 | -0.1317 | 2.0704 | 0.1000 | 0.7100 | |
| | 0.2634 | -0.1317 | -0.1317 | 2.4225 | 0.6000 | 0.7100 | |
| | 0.2634 | -0.1317 | -0.1317 | 2.7746 | 1.1000 | 0.7100 | |
| | 0.2634 | -0.1317 | -0.1317 | 3.1268 | 1.6000 | 0.7100 | |
| | 0.2634 | -0.1317 | -0.1317 | 3.4789 | 2.1000 | 0.7100 | |
| | 0.2634 | -0.1317 | -0.1317 | 3.8310 | 2.6000 | 0.7100 | |
| | 0.2634 | -0.1317 | -0.1317 | 4.1831 | 3.1000 | 0.7100 | |
| | 0.2634 | -0.1317 | -0.1317 | 4.5352 | 3.6000 | 0.7100 | |
| | 1.0273 | -0.8883 | -0.1390 | 2.0185 | 0.1000 | 2.7100 | |
| | 1.0273 | -0.8883 | -0.1390 | 2.1107 | 0.6000 | 2.7100 | |
| | 1.0273 | -0.8883 | -0.1390 | 2.2030 | 1.1000 | 2.7100 | |
| | 1.0273 | -0.8883 | -0.1390 | 2.2952 | 1.6000 | 2.7100 | |
| | 1.0273 | -0.8883 | -0.1390 | 2.3875 | 2.1000 | 2.7100 | |
| | 1.0273 | -0.8883 | -0.1390 | 2.4797 | 2.6000 | 2.7100 | |
| | 1.0273 | -0.8883 | -0.1390 | 2.5720 | 3.1000 | 2.7100 | Existing condition is $\beta > 1/\sqrt{2}$ and $\forall\lambda$. |
| | 1.0273 | -0.8883 | -0.1390 | 2.6642 | 3.6000 | 2.7100 | |
| | 1.0368 | -0.8208 | -0.2160 | 2.0226 | 0.1000 | 2.2100 | |
| | 1.0368 | -0.8208 | -0.2160 | 2.1357 | 0.6000 | 2.2100 | |
| *E* | 1.0368 | -0.8208 | -0.2160 | 2.2489 | 1.1000 | 2.2100 | Unstable point. |
| | 1.0368 | -0.8208 | -0.2160 | 2.3620 | 1.6000 | 2.2100 | |
| | 1.0368 | -0.8208 | -0.2160 | 2.4751 | 2.1000 | 2.2100 | |
| | 1.0368 | -0.8208 | -0.2160 | 2.5882 | 2.6000 | 2.2100 | Acceleration phase never occurs. |
| | 1.0368 | -0.8208 | -0.2160 | 2.7014 | 3.1000 | 2.2100 | |
| | 1.0368 | -0.8208 | -0.2160 | 2.8145 | 3.6000 | 2.2100 | |
| | 1.0437 | -0.5219 | -0.5219 | 2.0413 | 0.1000 | 1.2100 | |
| | 1.0437 | -0.5219 | -0.5219 | 2.2479 | 0.6000 | 1.2100 | |
| | 1.0437 | -0.5219 | -0.5219 | 2.4545 | 1.1000 | 1.2100 | |
| | 1.0437 | -0.5219 | -0.5219 | 2.6612 | 1.6000 | 1.2100 | |
| | 1.0437 | -0.5219 | -0.5219 | 2.8678 | 2.1000 | 1.2100 | |
| | 1.0437 | -0.5219 | -0.5219 | 3.0744 | 2.6000 | 1.2100 | |
| | 1.0437 | -0.5219 | -0.5219 | 3.2810 | 3.1000 | 1.2100 | |
| | 1.0437 | -0.5219 | -0.5219 | 3.4876 | 3.6000 | 1.2100 | |
| | 1.0485 | -0.5910 | -0.4575 | 2.0292 | 0.1000 | 1.7100 | |
| | 1.0485 | -0.5910 | -0.4575 | 2.1754 | 0.6000 | 1.7100 | |
| | 1.0485 | -0.5910 | -0.4575 | 2.3216 | 1.1000 | 1.7100 | |
| | 1.0485 | -0.5910 | -0.4575 | 2.4678 | 1.6000 | 1.7100 | |
| | 1.0485 | -0.5910 | -0.4575 | 2.6140 | 2.1000 | 1.7100 | |
| | 1.0485 | -0.5910 | -0.4575 | 2.7602 | 2.6000 | 1.7100 | |
| | 1.0485 | -0.5910 | -0.4575 | 2.9064 | 3.1000 | 1.7100 | |

FIGURE 1: Two and three dimensional projections of the phase-space trajectories for $\beta = 2.51$, $\lambda = 3.1$, $\lambda = 3.6$ and $\lambda = 4.1$. All plots begin from the critical point $A = (0.21, 0.70, 0.45, 0)$ being a stable attractor.

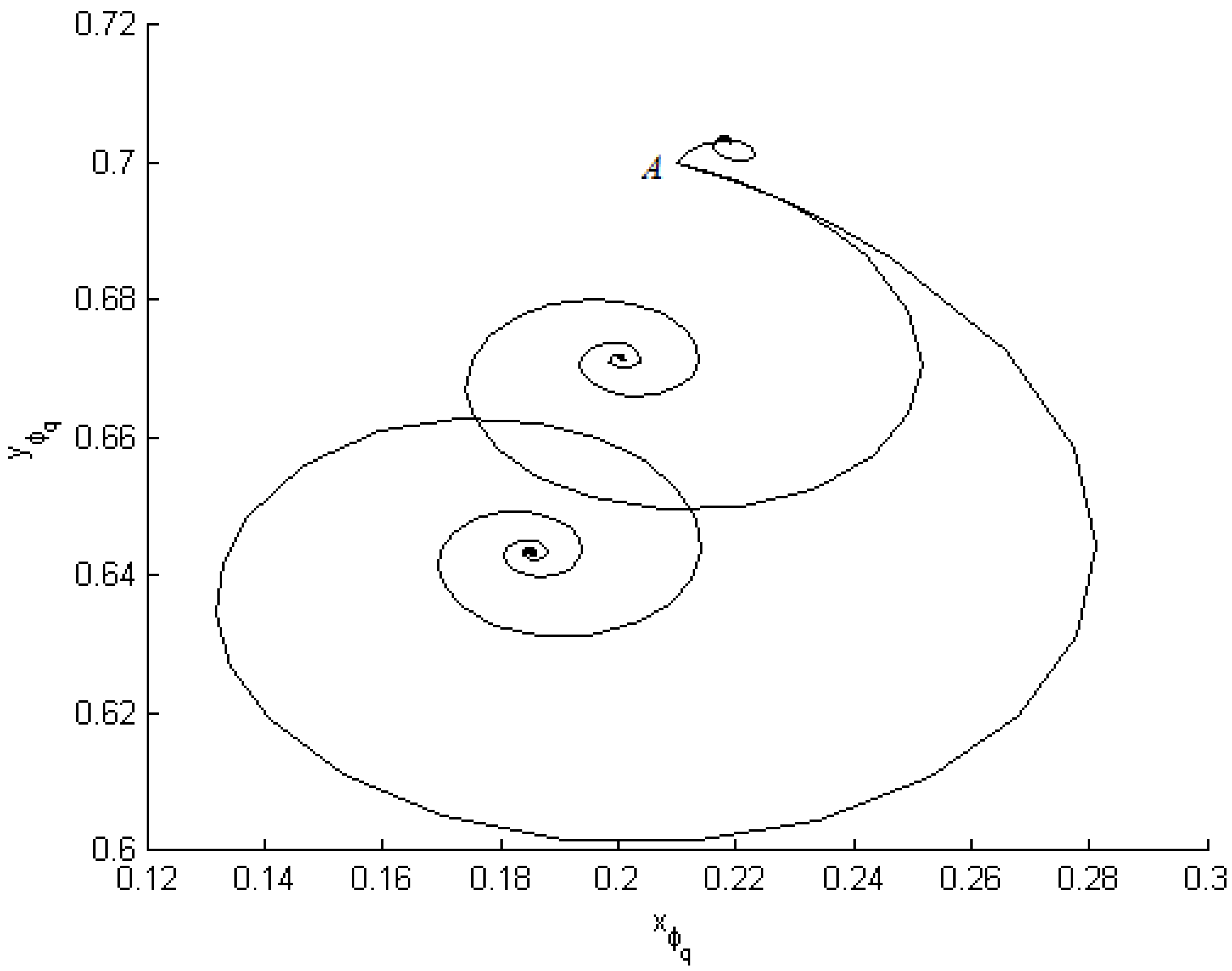

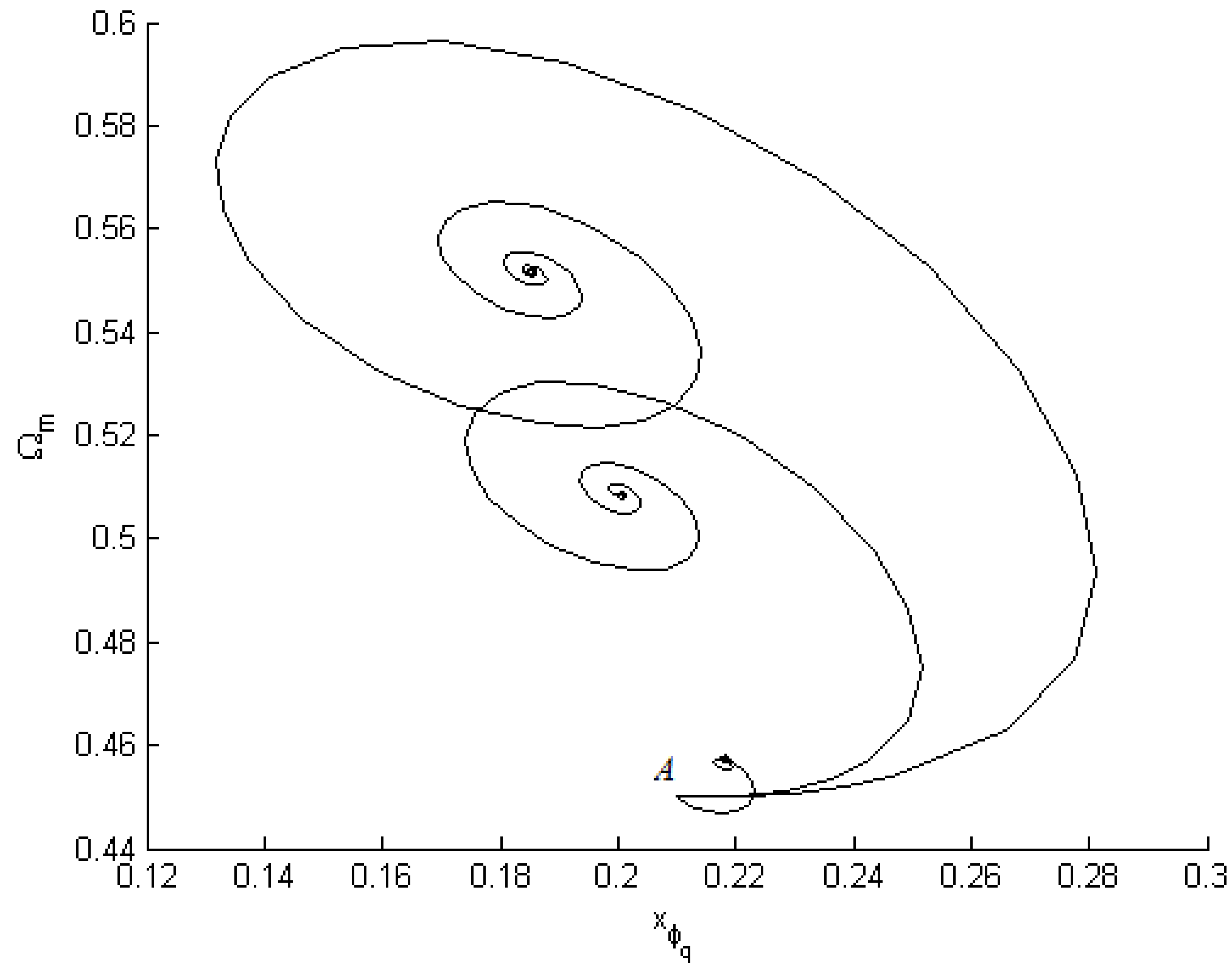

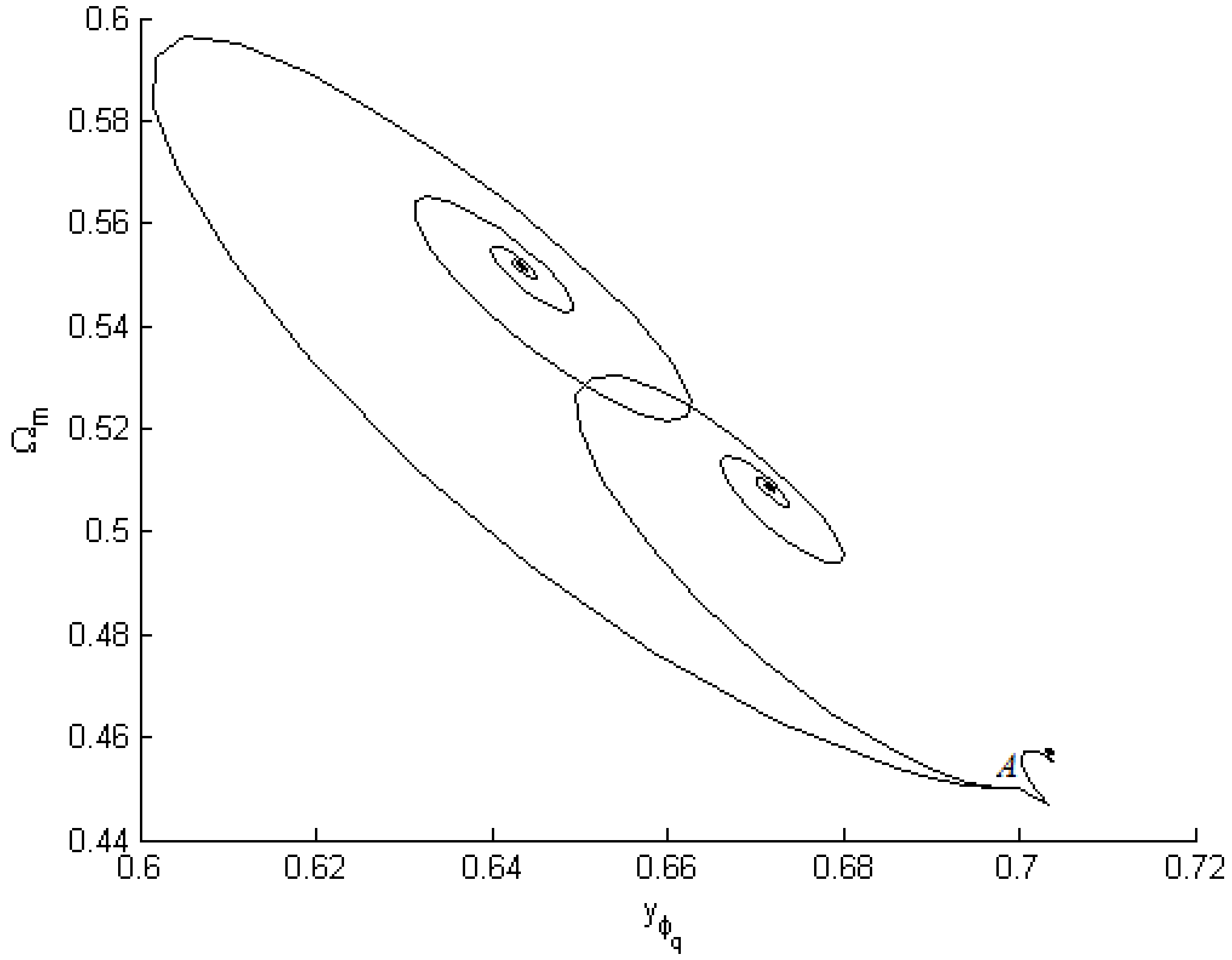

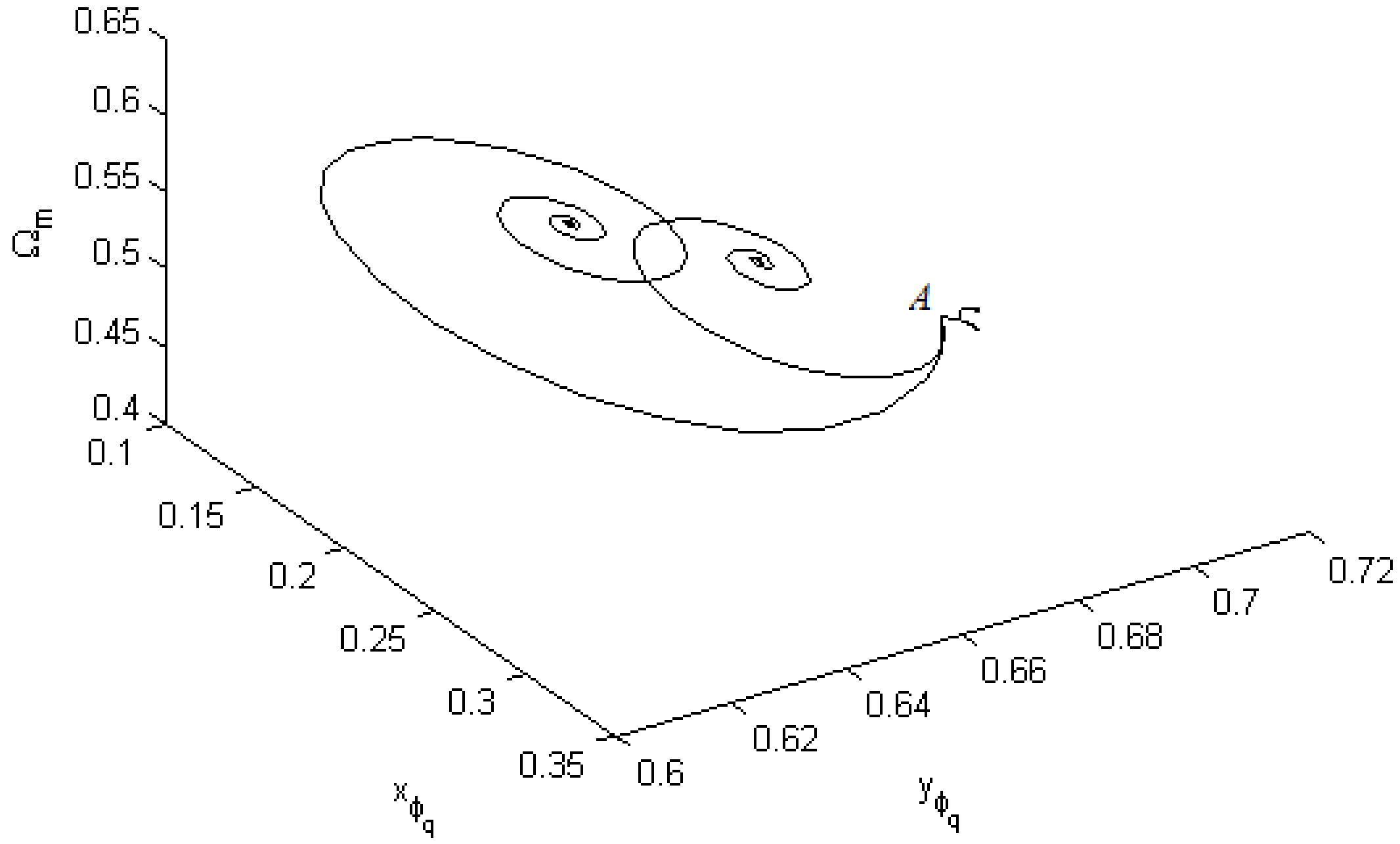

FIGURE 2: Two and three dimensional projections of the phase-space trajectories for $\beta = 2.51$, $\lambda = 3.1$, $\lambda = 3.6$ and $\lambda = 4.1$. All plots begin from the critical point $B = (0.21, -0.70, 0.45, 0)$ being a stable attractor.

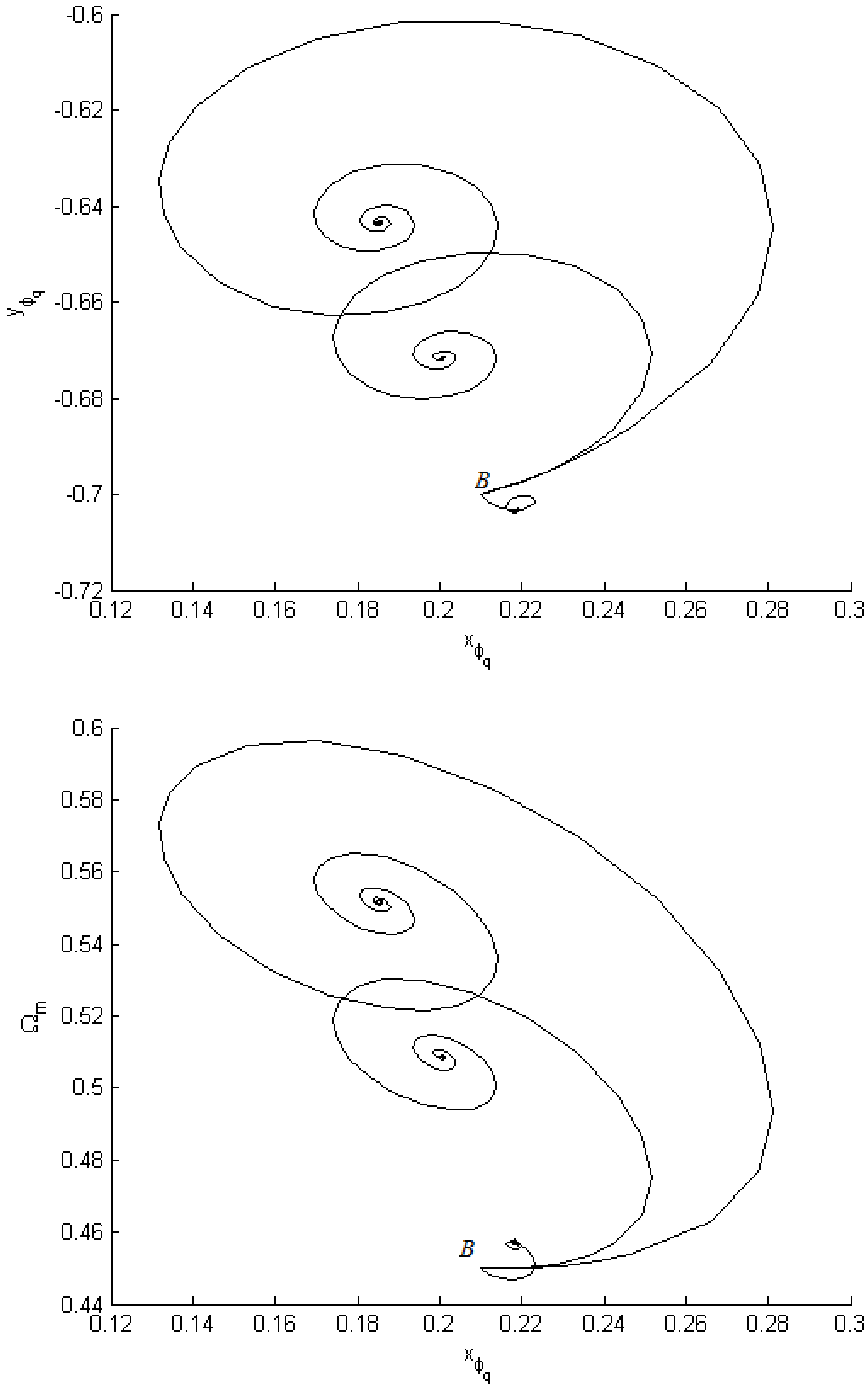

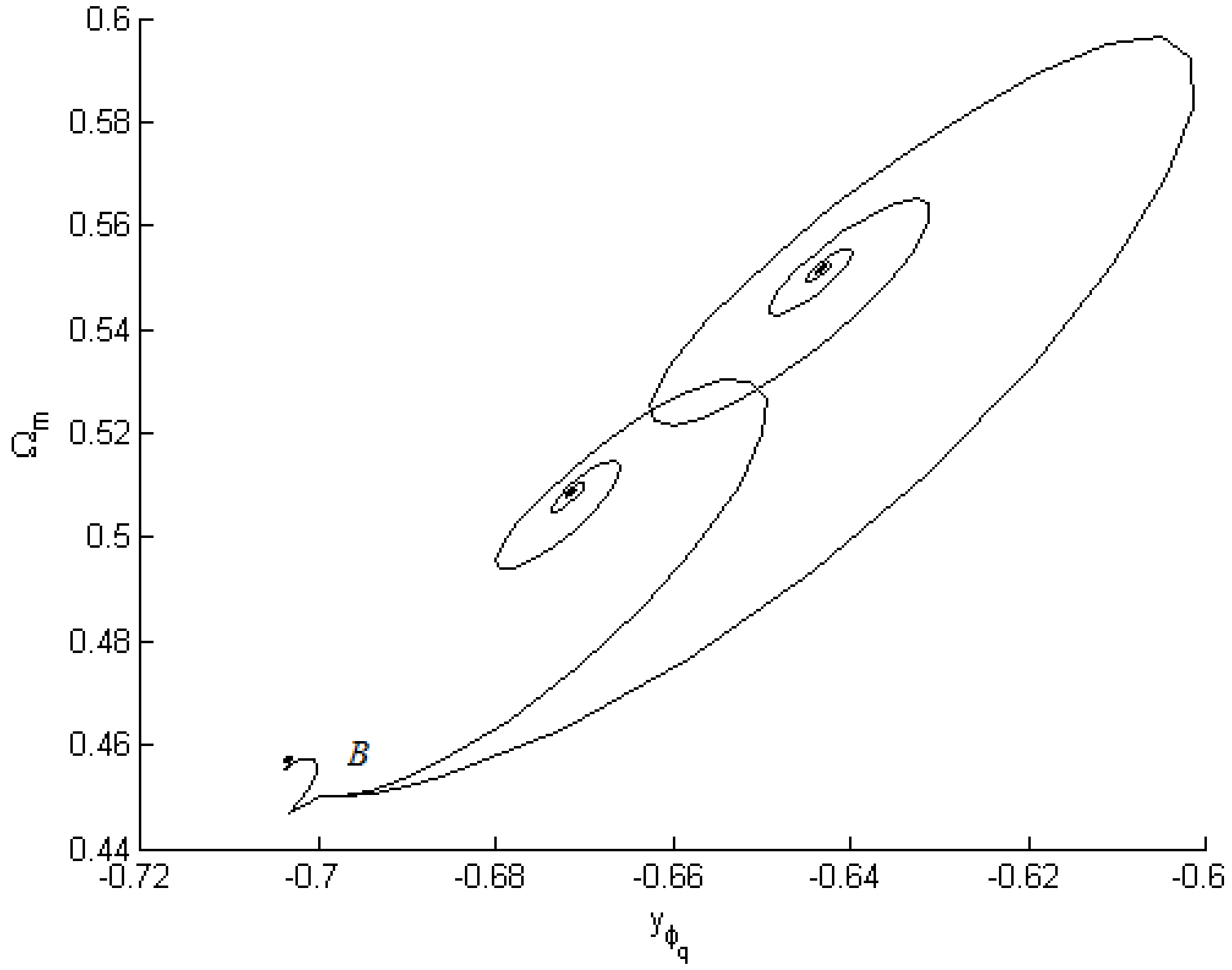

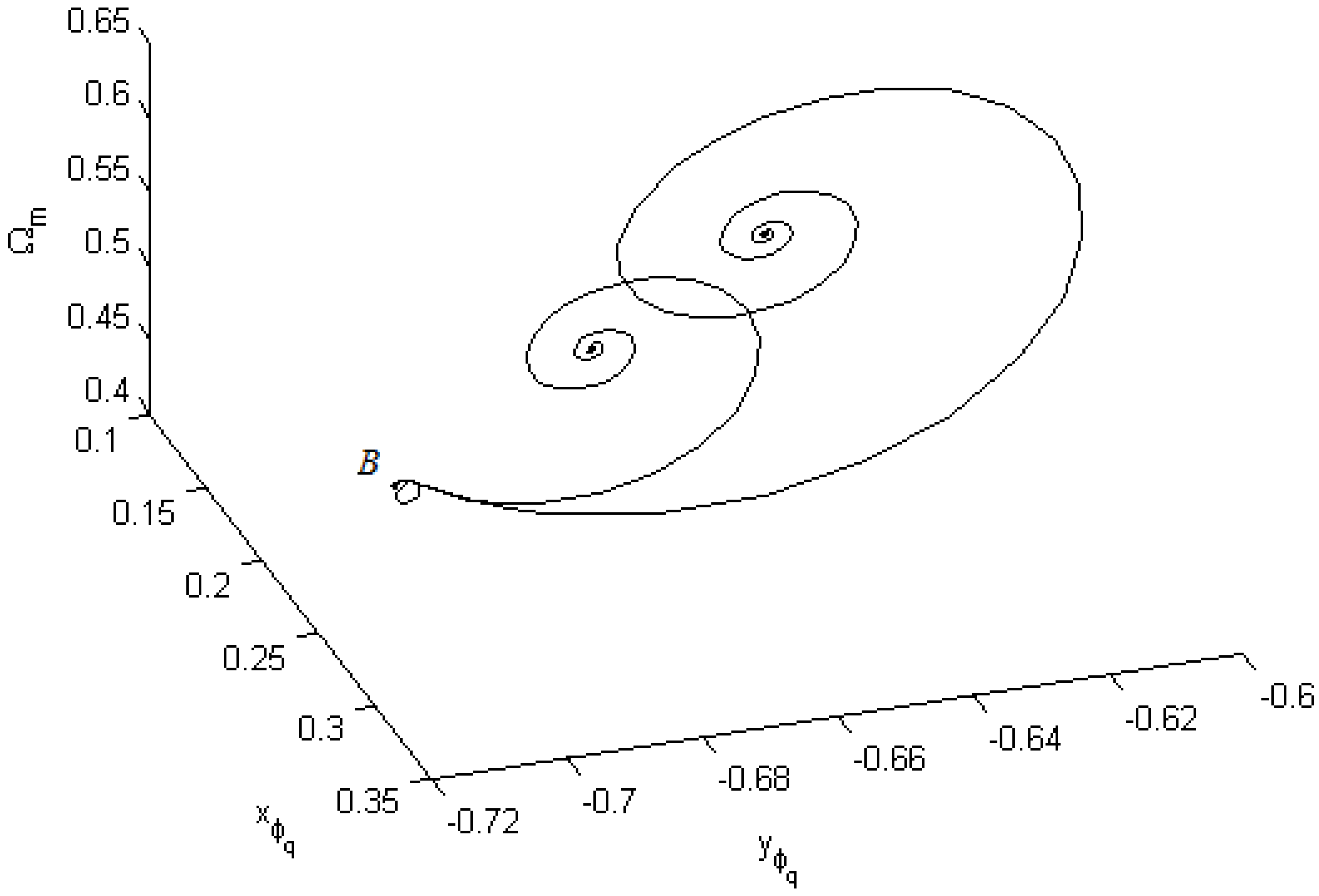

FIGURE 3: Two dimensional projections of the phase-space trajectories for $\beta = 1.5$, $\lambda = 0.001$, $\lambda = 0.5$ and $\lambda = 1.1$. All plots begin from the critical point $C = (0.48, 0.87, 0, 0)$ being a stable attractor.

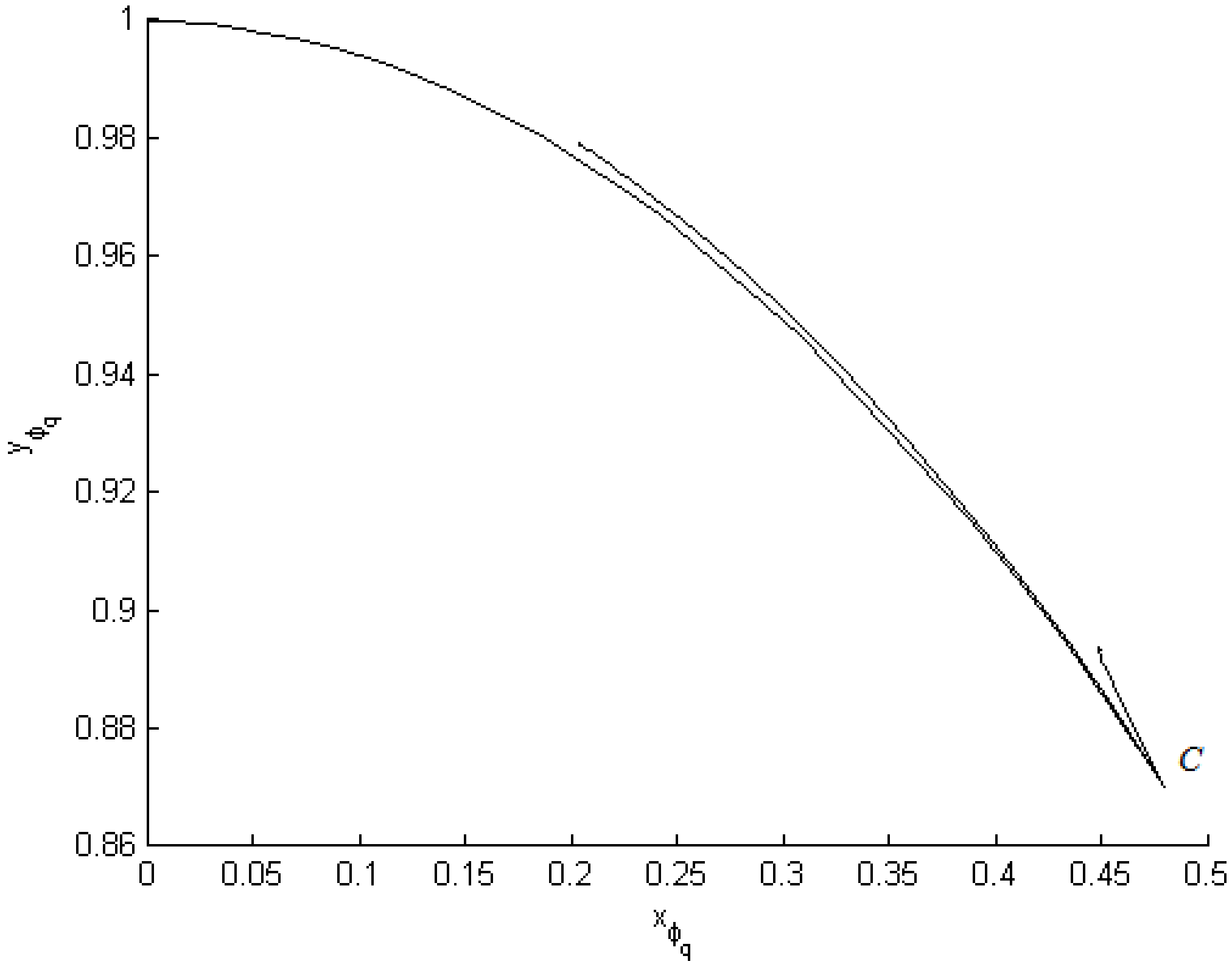

FIGURE 4: Two and three dimensional projections of the phase-space trajectories for $\beta = 1.5$, $\lambda = 3$, $\lambda = 4$ and $\lambda = 6$. All plots begin from the critical point $D = (0.54, 0.38, 0, 0.55)$ being an unstable solution.

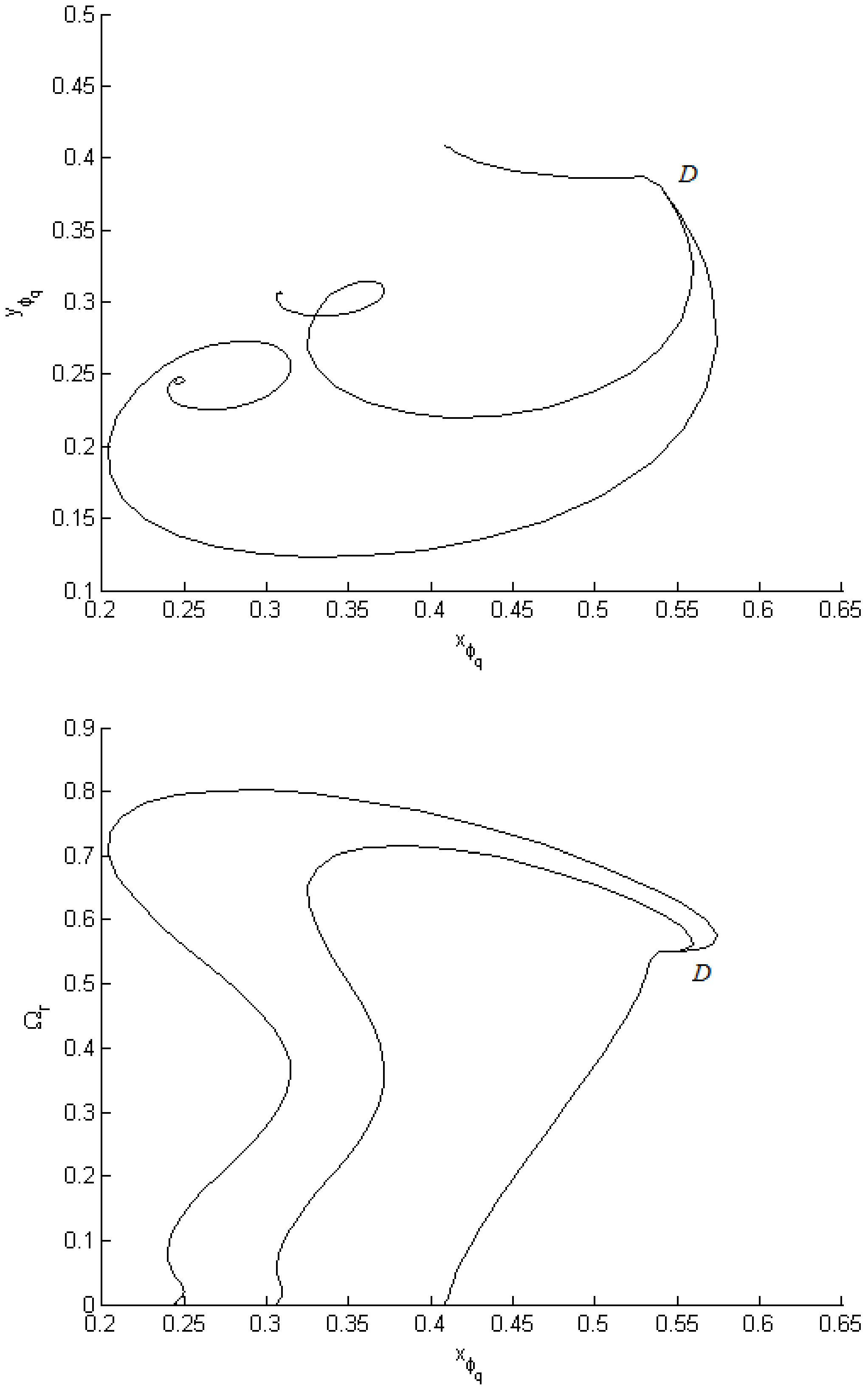

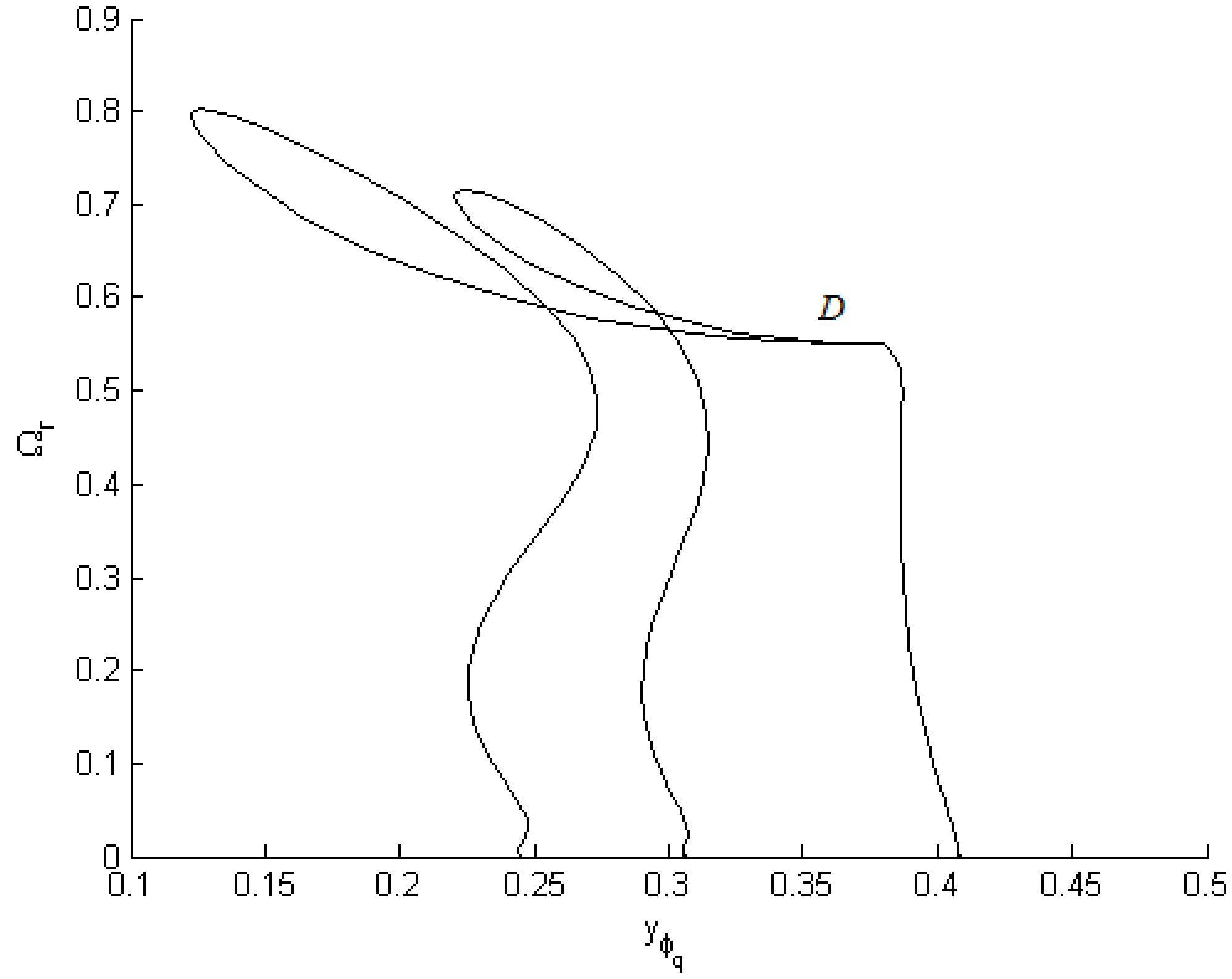

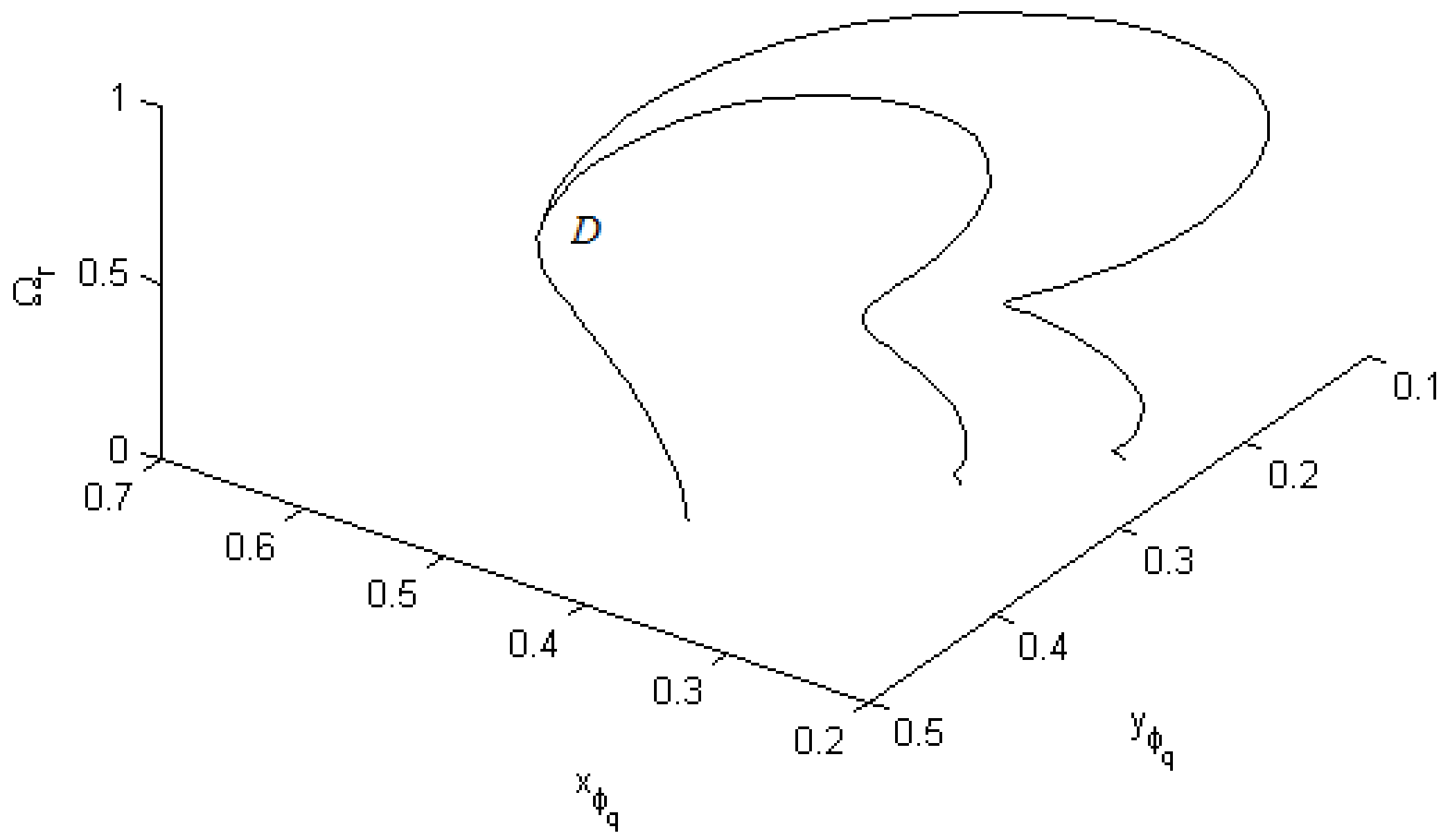

FIGURE 5: Two and three dimensional projections of the phase-space trajectories for $\lambda = 1$, $\beta = 1.7$, $\beta = 2.6$ and $\beta = 3.5$. All plots begin from the critical point $E = (-0.24, 0, 0.11, 0.82)$ being an unstable solution.

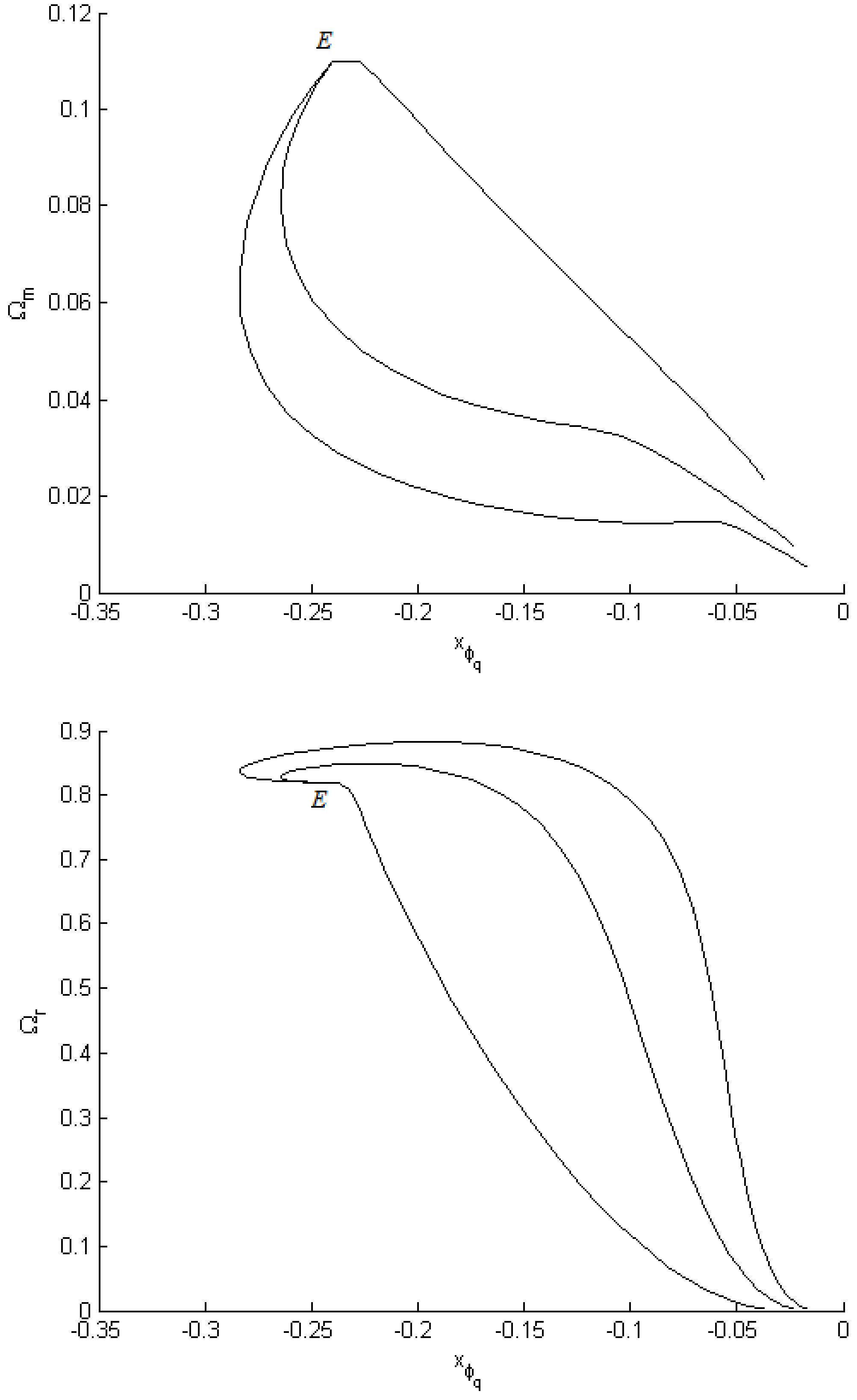

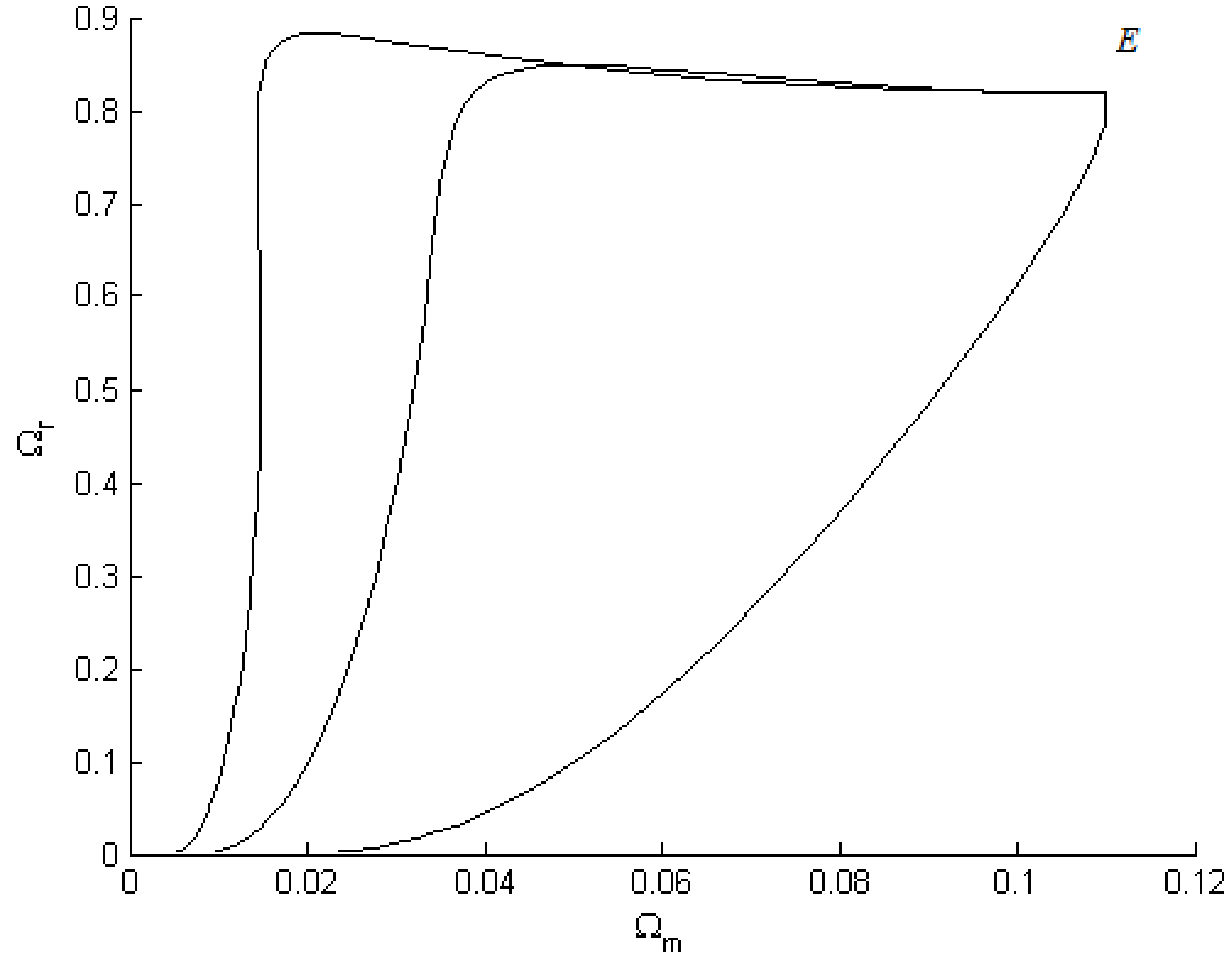

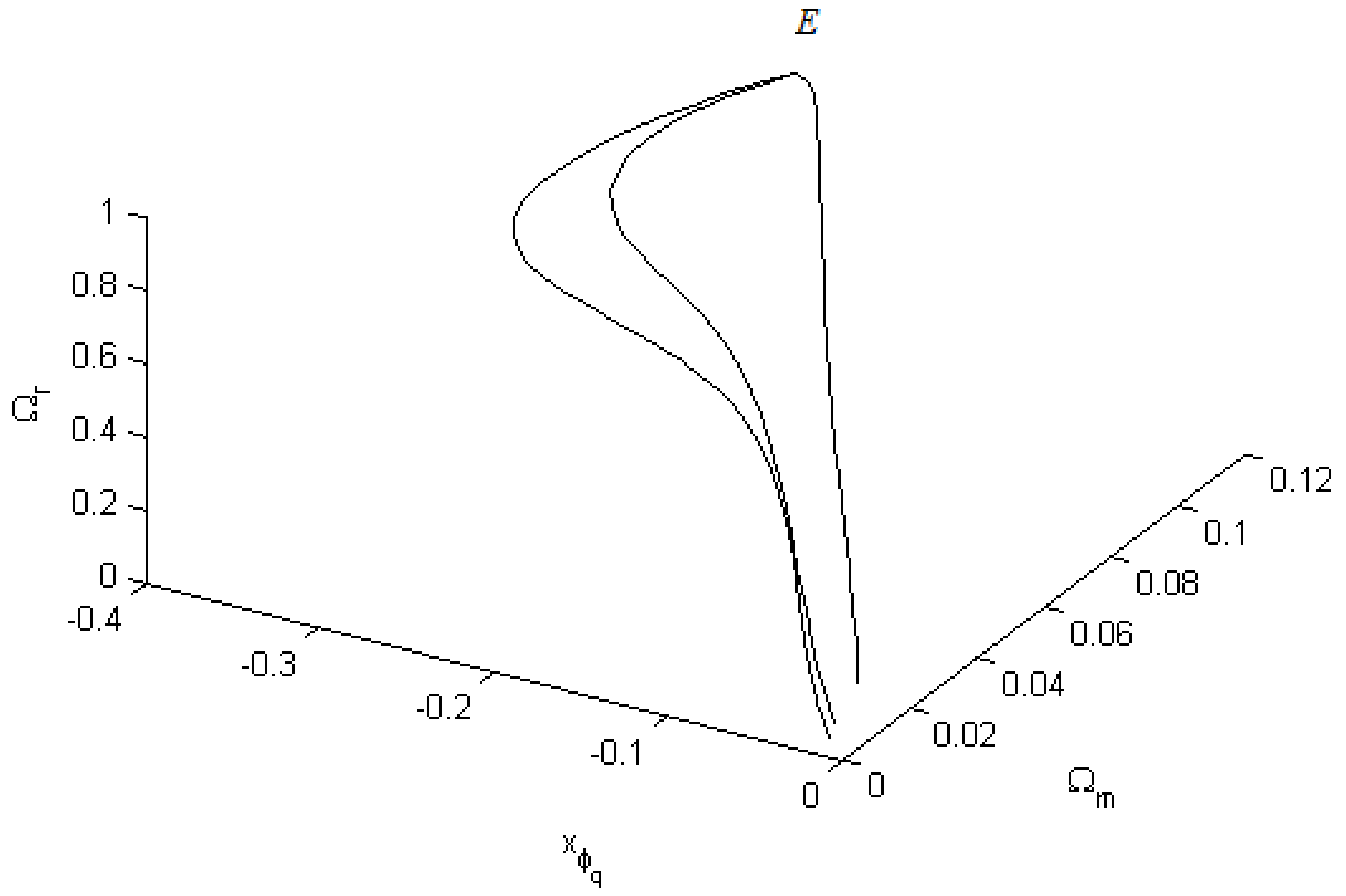

FIGURE 6: $q$-deformed energy density for various values of $N$ and $q$, in terms of standard energy density.

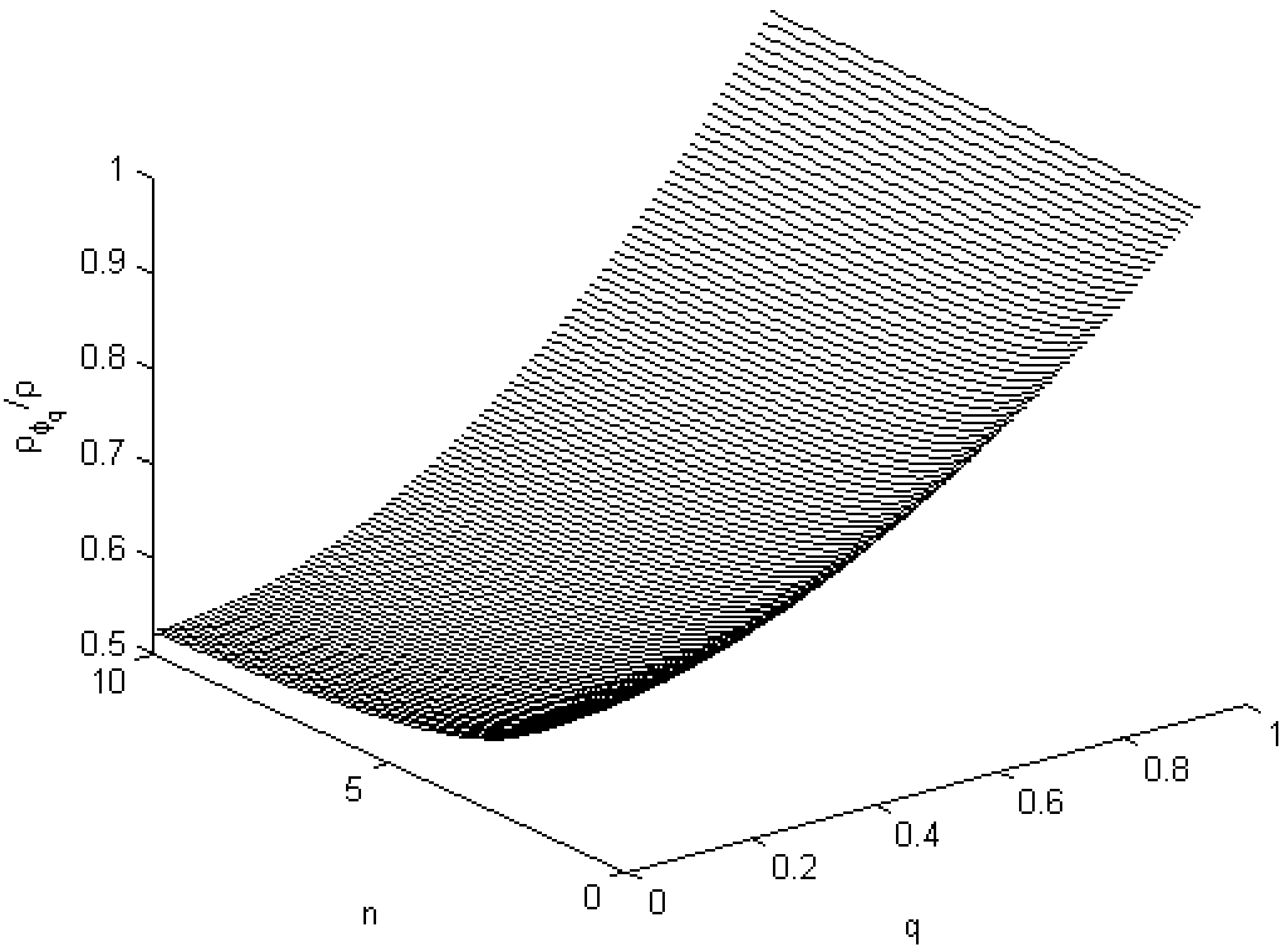

FIGURE 7: $q$-deformed pressure for various values of $N$ and $q$, in terms of standard pressure.

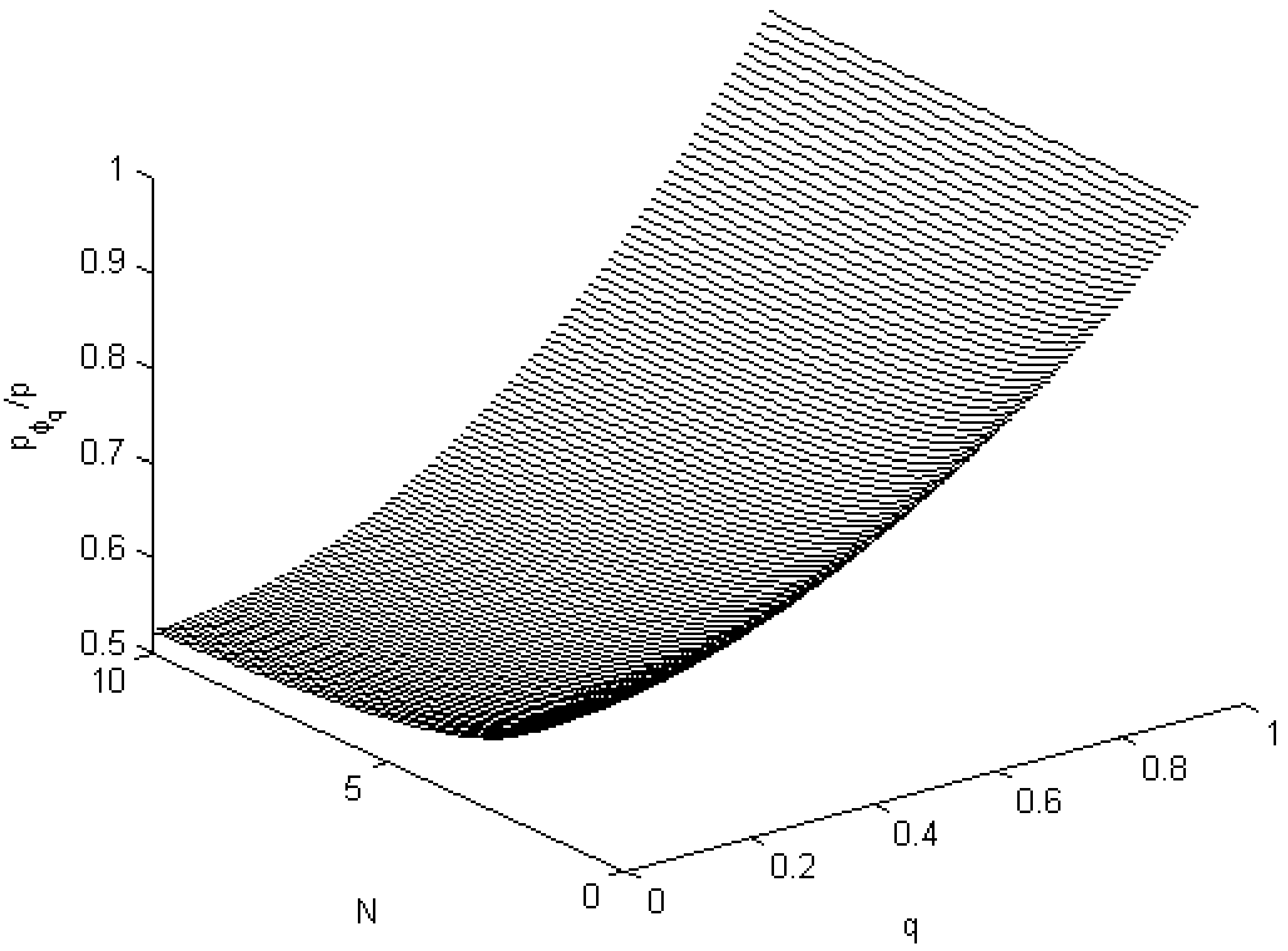

FIGURE 8: $q$-deformed scalar field for various values of $N$ and $q$, in terms of standard scalar field.

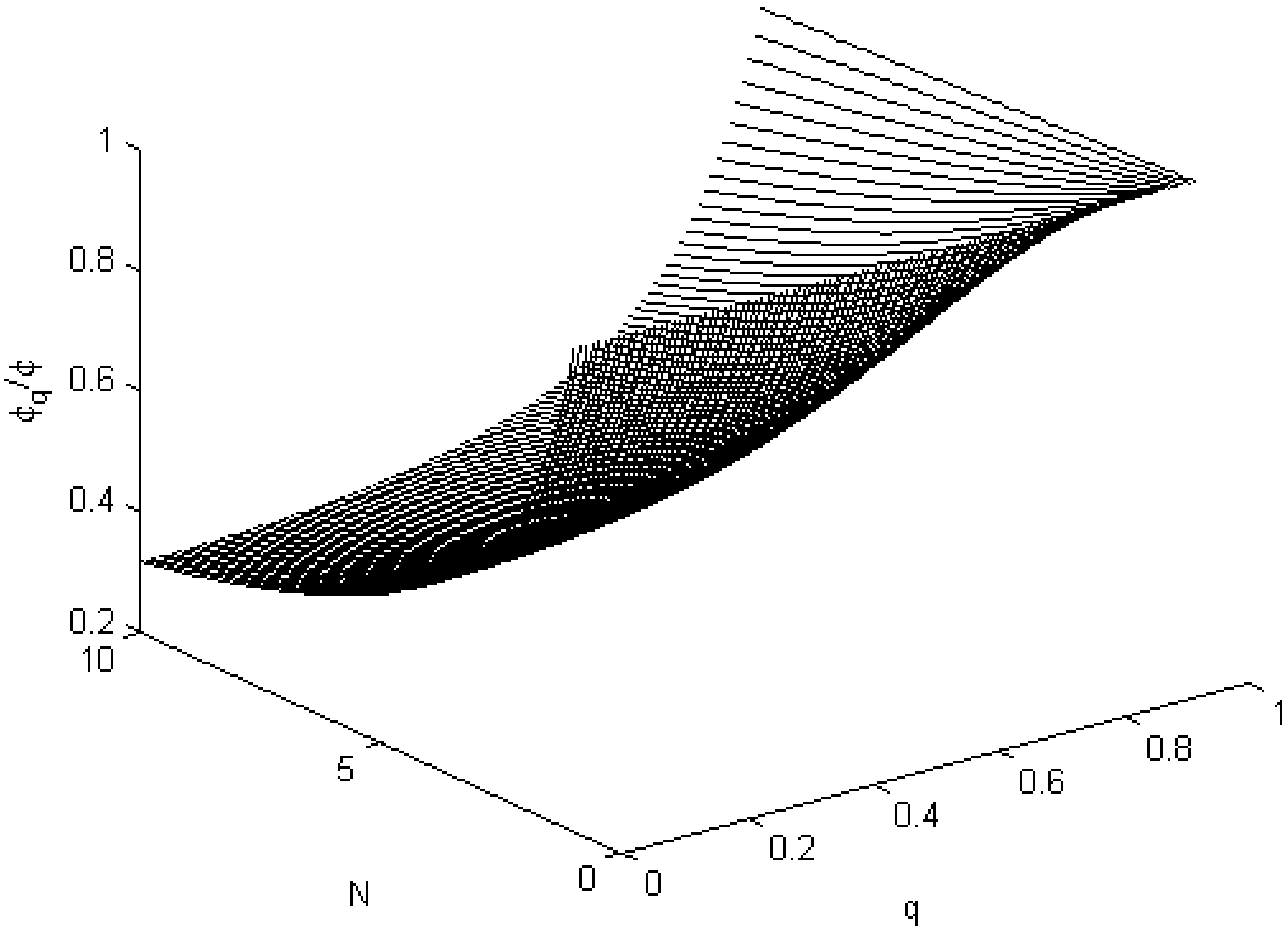

FIGURE 9. Effect of $q$ on the accelerated expansion behavior with negative dark energy pressure.

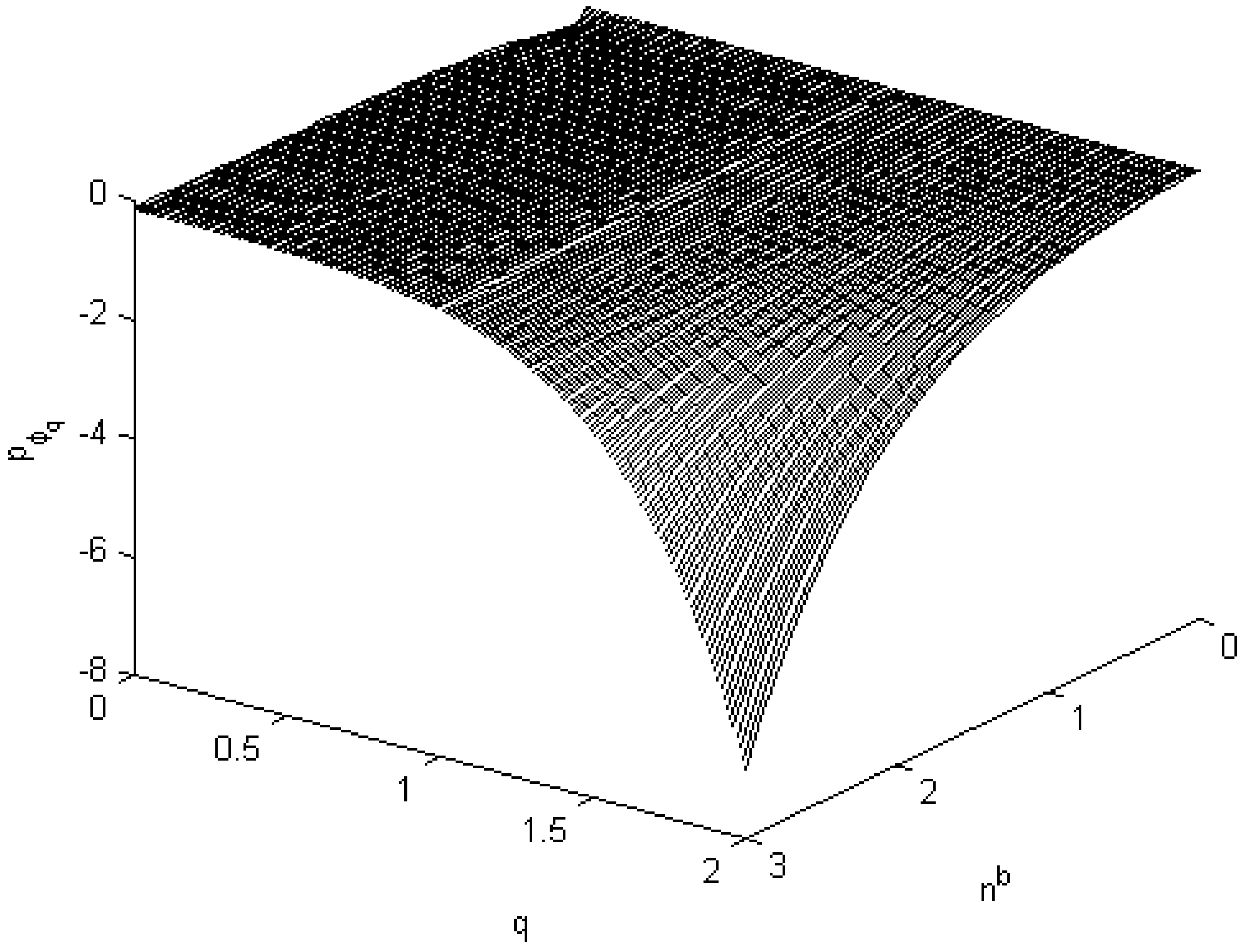

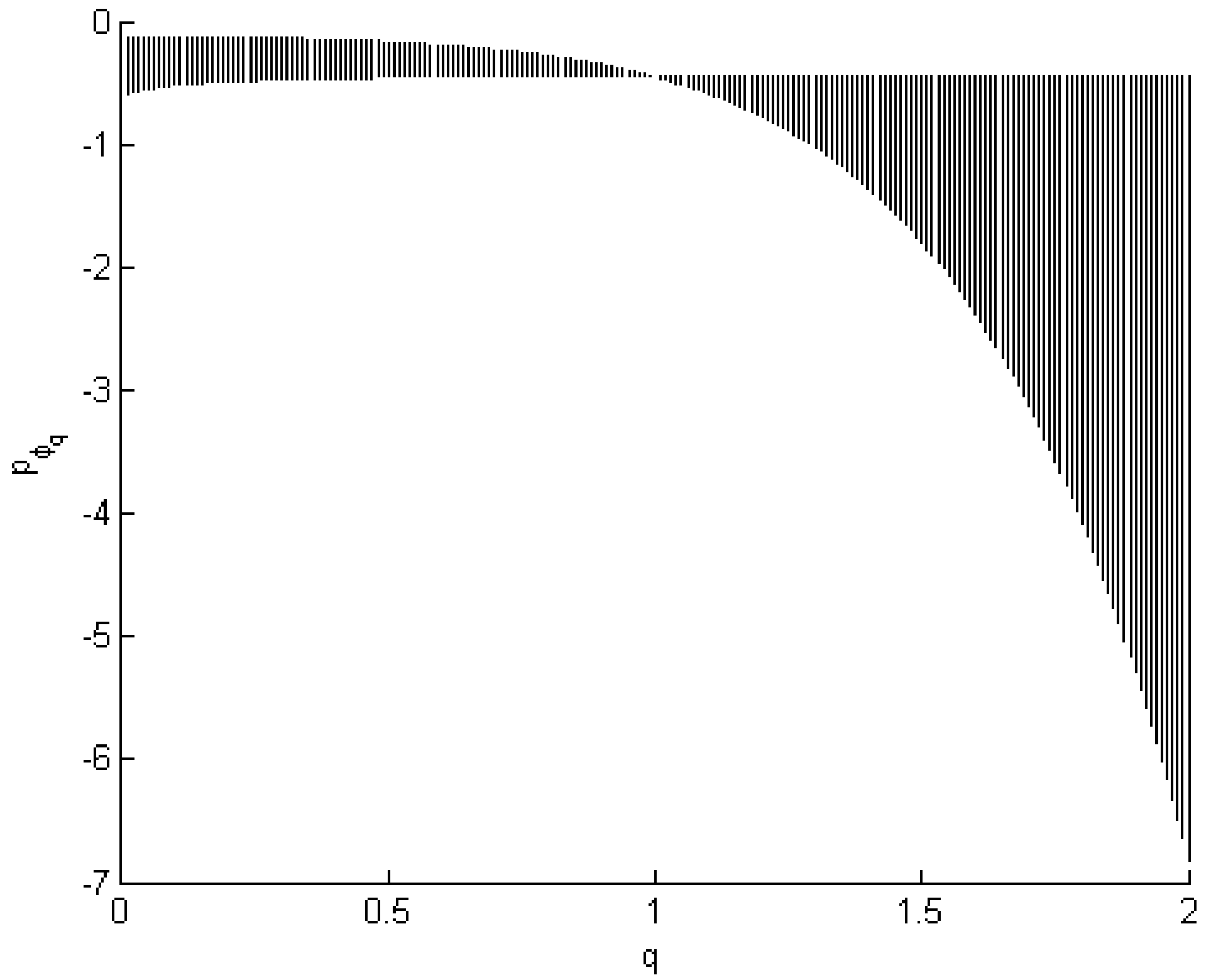